\documentclass[12pt,article,fleqn]{article}
\usepackage{amsthm}
\usepackage{amsmath,amsfonts,amssymb}
\usepackage{hyperref}
\usepackage{graphicx,wrapfig}
\usepackage{float}
\usepackage[utf8]{inputenc}
\usepackage[font={it},labelfont=bf]{caption} %make captions italic
\usepackage{longtable}

\usepackage{enumitem}
\usepackage{xcolor}

\theoremstyle{definition}
\theoremstyle{plain}
\newtheorem{assumption}{Assumption}%[section]
\newtheorem{theorem}{Theorem}%[section]
\newtheorem{corollary}{Corollary}%[theorem]
\newtheorem{remark}{Remark} % to suppress numbers: put * after \newtheorem

\usepackage[boxed]{algorithm2e}
\usepackage{multirow}
\usepackage{verbatim}

\usepackage{dcolumn}

\usepackage[backend=bibtex,citestyle=authoryear,bibstyle=authoryear,maxbibnames=9, maxcitenames=2,natbib]{biblatex}%citestyle=apa, the et al truncation only works with authoryear style

\renewbibmacro{in:}{}

\newcommand{\W}{W} % used to be "\mathbf{W}", I simply changed this and the following commands when we decided to use non-bold W and w
\newcommand{\w}{w}

\newcommand{\indept}{\mathrel{\perp\mspace{-10mu}\perp}}

\newcommand{\norm}[1]{\left\lVert#1\right\rVert}

 \author{{{  Christoph Breunig}}\thanks{University of Bonn, 53113 Bonn, Germany. Email: \url{	cbreunig@uni-bonn.de}}
  \and{{Patrick Burauel}}\thanks{California Institute of Technology, Pasadena, CA 91125,  USA. Email: \url{pburauel@caltech.edu}}}
\date{\today }

 \title{
Testability of Reverse Causality Without \\ Exogenous Variation}

\begin{document}
	
	\maketitle

	\begin{abstract}
%			\normalsize

This paper shows that testability of reverse causality is possible even in the absence of exogenous variation, such as in the form of instrumental variables. Instead of relying on exogenous variation, we achieve testability by imposing relatively weak model restrictions and exploiting that a dependence of residual and purported cause is informative about the causal direction.
Our main assumption is that the true functional relationship is nonlinear and that error terms are additively separable. We extend previous results by incorporating control variables and allowing heteroskedastic errors. We build on reproducing kernel Hilbert space (RKHS) embeddings of probability distributions to test conditional independence and demonstrate the efficacy in detecting the causal direction in both Monte Carlo simulations and an application to German survey data.
% We can infer the causal direction between income and work experience (proxied by age) without relying on exogeneous variation.
%		Due to the nature of the problem, an asymptotic distribution under the null hypothesis cannot be derived.
%We demonstrate the usefulness of our test in Monte Carlo simulations and an empirical illustration.\pcomment{Christoph, bitte nochmal checken, hier war noch von "testing reverse causality" die Rede, das habe ich geaendert}
%		Critical values cannot be derived due to the nature of the problem. 
		
%		, as one type of endogeneity, of a single regressor without requiring instruments. Key assumptions to establish testability are a nonlinear 
%		
%		The fundamental assumption is that the data generating process can be represented by a regression model with additively separable error term and a nonlinear relation between cause and effect.
		%We move beyond the mean independence assumption and require the error to be independent of the regressor, which allows us to make inference on the causal direction between the variables at hand.  We leverage advances on kernel-based testing of conditional independence of random variables. 
	\end{abstract}


%\spacingset{1.8} % DON'T change the spacing!
\section{Introduction}
%
%\begin{itemize}
%\item Stucture Introduction in paragraphs
%\begin{enumerate}
%\item Relevance of research question
%\item Central contribution of our paper
%\item Key difference between our paper and existing literature
%\item Limitations of the procedure in terms of critical values
%\item Implementation via RKHS
%\item Finite samples
%\end{enumerate}
%\end{itemize}

Endogeneity is a central problem in econometric models which potentially invalidates estimates of causal effects. %Existing tests of endogeneity often require that a potential solution in the form of instruments is available. In many situations, such exogenous variables might violate exclusion restrictions or may not be available. 
Reverse causality, where the dependent variable causes the independent variable,  is one source of such endogeneity. The aim of this paper is to show that reverse causality is testable under mild assumptions and without relying on exogenous variation. %We show that that testability can be achieved under mild model restrictions, which appear plausible in many microeconomic applications.
% Yet, as we show detailed in simulations, models that are close to linear have limited power to detect each model direction. 
We build on \textcite{hoyer09anm} who provide a link between nonlinear model structure and causality, namely,  that a nonlinear relation between cause and effect leads to observable signals about causal direction using observational data. While their theoretical results are striking, their results assume homoskedastic errors and do not generalize to settings with additional control variables. 
We show that this assumption can be relaxed to allow for heteroskedasticty with respect to additional control variables, as is commonly the case in econometric applications. 
Our primary conditions for achieving testability are twofold: First, we necessitate a nonlinear relationship between the dependent variable and the regressor (with no restrictions on how controls are incorporated into the model). Second, we impose additively separable errors.
% We emphasize that our testability results hold in the presence of heteroskedastic errors w.r.t. additional control variables.

Following \cite{hoyer09anm}, we also demonstrate how our testability result can be applied in empirical practice. Specifically, we show that identification of the causal direction is equivalent to a conditional independence test of covariates and error terms given control variables. We make use of conditional independence tests based on kernel mean embeddings, i.e., maps of probability distributions into reproducing kernel Hilbert spaces (RKHS) \parencite[see][for a survey]{muandetetal16kernel}. Intuitively, this corresponds to approximating conditional distributions with unconditional ones by weighting with an appropriate kernel, and evaluating their covariance in an RKHS. 
The method can detect nonlinear dependencies.

We consider two formal applications of our testability result. First, we explore testing for causal direction based on conditional independence. As already indicated in the related literature, achieving exact size control can be challenging within this framework, and we provide a detailed discussion of this issue below.  Second, we conduct causal discovery,  where we remain agnostic about the causal direction and compare test statistics for two rivaling models to gain insight into which one represents the true causal structure \parencite[see][]{peters14}.

In Monte Carlo simulations, we investigate the power of our approach to detect reverse causality.  We see that the degree of nonlinearity increases the power to detect reverse causation. The procedure has surprisingly high accuracy in detecting the true causal direction even under moderate form of nonlinearities and can be powerful even in  linear models under restrictions on the distribution of errors terms, which is a result described by \citep{shimizu06lingam}. 
%We show that our procedure has high accuracy in detecting the true causal direction in simulated data. 
Furthermore, we provide an empirical illustration  using data from the German Survey of Income and Expenditure. We show that our algorithm can infer from purely observational data that work experience is a causal driver for income, not vice versa. Substantively, this is not a surprising result; however, the fact that it can be inferred without exogenous variation is.

\paragraph{Related literature}\label{sec_related_lit}
%There is a vast literature on causal structure learning \parencite{eci17}, which develops without receiving much attention from the economics profession. Notably, \textcite{imbens19po_vs_dag} argues that this research ``has not had as much impact in economics as it should have'' (p. 1). Our paper aims to be a first remedy to that situation.
%We build on \textcite{hoyer09anm}, who establish that the causal direction between two variables that are nonlinearly related is identifiable. The main contribution of this paper is to extend the framework of \textcite{hoyer09anm} to the case where there is heteroskedasticity w.r.t. additional covariates and clarify the usefulness of this approach in economic applications.

Our test rests on the idea that $ X $ causing $ Y $ implies an independence between the error of a regression of $ Y $ on $ X $, and $ X $. 
%This idea has precursors in the literature. In particular,
This idea goes back to \textcite{engleetal83}, who propose a definition of an exogenous relation in terms of conditional densities. In particular, they argue that, if a joint probability density of two random variables $ Y $ and $ X $ factorizes as $ f(Y,X) = f(Y|X)f(X)  $ and the conditional density $ f(Y|X) $ is invariant to changes in the marginal density $ f(X) $, then $ X $ is called ``super exogenous" (p. 278). Statistical tests for the notion of ``super exogeneity" are proposed by \textcite[][]{faverohendry92,englehendry93,hendrysantos10}. These tests rely on analyzing to what extent parameter values are sensitive to exogenous interventions on the purported cause. Thus, their results specifically rely on exogenous variation (e.g. in the form of instrumental variables) whereas the approach at hand does not require such variation. 
%We interpret the invoked invariance to changes in terms of independence between the true error and the covariate and derive testable implications.

The problem of identifying causal structure from non-experimental data is receiving considerable attention in the causal machine learning literature \parencite[for comprehensive overviews see][]{mooijetal16,eci17,scholkopf2021toward}. In its bivariate form, the problem is concerned with deciding whether a variable $ X $ is causing $ Y $ or vice versa solely based on a non-experimental joint probability distribution of the two variables. Without making any assumptions regarding the true underlying data-generating process, no identification is possible.\footnote{Previous work shows that the causal direction cannot be identified without making further assumptions. \textcite[][Proposition~2.6]{peters12} proves that for every joint distribution of two variables, $ X $ and $ Y $, there is a model $Y = h(X,\varepsilon), \; \text{with} \; X\indept\varepsilon $ with $ h $ a measurable function and $ \varepsilon $ a real-valued noise variable. The roles of $ X $ and $ Y $ can be easily interchanged showing that the joint distribution itself does not identify the causal direction in this most general form.} \textcite{shimizu06lingam} show that non-Gaussianity of observed variables leads to identifiability. Subsequently, \textcite{hoyer09anm} show that nonlinearity of $ h $ can play a similar role as regards the identifiability of the causal direction as non-Gaussianity. If the true model is of a nonlinear form, one can infer the causal direction without making any assumptions about the distribution of the error. %Thus, our work is closely related to this literature and, therefore, we agree with \textcite{imbens19po_vs_dag}, who argues that it ``has not had as much impact in economics as it should have'' (p. 1), and show how insights developed in the causal discovery community can be useful to solve problems in economics. 

\textcite{hoyer09anm} and \textcite{mooijetal16} discuss inference of the causal direction between \textit{two} random variables (cause and effect) from observational data. \textcite{peters14} constitutes a theoretical extension of these methods to more than two variables. The paper at hand falls between these two strands as it accounts for more than two variables, yet its primary concern is the causal directionality between a subset of just two of them. The remaining variables $\W$ serve as controls.

%To make these theoretical identifiability results operational, i.e. useful in applications with real data, we leverage novel conditional independence testing that rely on flexible mappings of probability distributions into reproducing kernel Hilbert spaces (RKHS), so-called kernel mean embeddings. 
In this paper, we rely on RKHSs and kernel mean embeddings, which are not widely used in the econometrics literature, albeit with notable exceptions.  \textcite{carrasco00continuum_gmm}  and \textcite{carrasco07} discuss the usefulness of RKHS theory given infinite number of moment conditions. 
%This seemingly rare situation might occur when the moment conditions can be expressed as a \textit{function}, i.e. a vector of infinite length.
% For instance, \textcite{carrasco00continuum_gmm} further generalize (already) generalized method of moments estimators to account for infinitely many moment conditions. To analyze the infinitely many moment conditions requires inverting a covariance operator. 
%Akin to the procedure described here, they show that the generalized inverse of such operator only exists in the RKHS.
 \textcite{singh19kernel_iv} study the use of kernel methods in the context of instrumental variable (IV) methods. They use kernel mean embeddings of the conditional distribution of the covariates given the instrument to propose a nonlinear extension of linear IV implementations. \textcite{zhang2020maximum} propose kernelized moment restrictions to estimate nonlinear IV estimators. \textcite{grunewalder12} analyze connections between kernel mean embeddings and vector-valued functions to analyze Markov decision processes. \textcite{flaxman15} use kernel mean embeddings to analyze who cast their votes for Obama in the 2012 US presidential election.

This paper is also related to a strand of the literature, which make use of exogenous variations to detect endogeneity of regressors. The idea to make use of instrumental variables to detect endogeneity was originally proposed by \textcite{hausman78specification}. More recently,  \textcite{blundell07exog} and \textcite{breunig15} provide exogeneity tests using instrumental variables for nonparametric models with additively separable errors, \textcite{feve18estimation} and \textcite{breunig18specification} for models with nonseparable errors.\\

The remainder of the paper is organised as follows. Section~\ref{sec:testability} establishes testability of reverse causality under nonlinear regression functions and introduces the RKHS test for independence. In Section~\ref{sec_mcs}, we analyze
finite sample power of our RKHS procedure in a Monte Carlo simulation study. Section~\ref{sec_asymmetry} provides an application of our method to empirical data. 
Appendix~\ref{app_proof} provides a proof of our main testability result. 
Appendix~\ref{app_rkhs} gives a review of the construction of RKHS. 
Appendix~\ref{app_more_results} contains additional Monte Carlo simulation results. 

\section{Testing Reverse Causality}\label{sec:testability}

We show how to test for reverse causality between two variables $ X $ and $ Y $ in the presence of additional covariates $ \W $.
% where $ \W $ need not be independent of the regression error. 
First, we introduce the model, discuss how the model specification relates to the existing causal discovery literature, and derive testable implications. Second, we present the conditional independence test that is a central component of the test. Third, we present the implementation of the test.

\subsection{Model and Assumptions}

% In the following, we discuss the assumptions we make that enable us to identify the causal direction.

Consider a model where observable continuous scalar variable $ X $ causes observable scalar variable $ Y $ in the presence of the vector of covariates $\W$ (which we refer to in the following as the \textit{model}):
\begin{equation}\label{eq_w_causal}
Y=h(X,\W) + U \quad \text{where}\quad U = \sigma(\W) \; \varepsilon\;  \; \text{and}\;  \; \varepsilon \indept (X,\W) \
%\quad\quad \text{(\textit{model})}
\end{equation}
where  $\varepsilon$ are unobservable variables and $ \sigma(\cdot)$ some strictly positive function. In addition, we assume $E[\varepsilon]=0$ without loss of generality.

Note that the error $ U $ is additively separable. The additive separability of $ U $ precludes the dependence of marginal effects on unobservables except through a dependence via $W$. The model allows for heteroskedasticity of the error term $U$ with respect to the control variables $ \W $. In particular, model equation \eqref{eq_w_causal} implies $  U \indept X|\W  $, which is also known as conditional exogeneity \parencite[see][]{whitechalak2010condtlexo}. It corresponds to the unconfoundedness assumption in the treatment effects literature \parencite{imbensrubin15} and is also closely related to the special regressor assumption \parencite[see][for an overview]{lewbel14}. 
%\ccomment{The average marginal treatment effect is identified by $\mathbb E[\partial_x h(X,W)]$}\pcomment{why did you include this sentence here?}.

The main idea of this paper is to study the conditions under which this model is distinguishable form reversed analog without relying on exogenous information. The \textit{reverse model}, where $ Y $ is causing $ X $, again  in the presence of the vector of covariates $\W$, is defined as
\begin{equation}\label{eq_w_backw1}
X=\widetilde{h}(Y,\W)+ \widetilde{U} \quad \text{where}\quad \widetilde{U} = \widetilde{\varepsilon} \; \widetilde{\sigma}(\W)\text{ and } \widetilde{\varepsilon} \indept (Y,\W) 
%\quad\quad \text{(\textit{reverse model})}
\end{equation}
where  $\widetilde{\varepsilon}$ are unobservable variables and $ \widetilde{\sigma}(\cdot)$ some strictly positive function.\\
%\ccomment{HERE MORE DISCUSSION: EVEN IN $h$  IS LINEAR THEN $Y=X+ U$ THEN THE REVERSE MODEL IS $X=Y-U$. IT IS NOT ALWAYS TRUE THAT WE CAN REWRITE THIS REVERSE MODEL AS $X=Y+\widetilde U$, RIGHT? CAN WE GIVE A SIMPLE EXAMPLE, WHERE IN THE LINEAR CASE WE CAN FIND A REVERSE MODEL?
%\begin{example}
%...
%\end{example}}\pcomment{I moved this discussion down[comment can be deleted]}

We denote the probability density function of a random vector $V$ by $f_V$ and make the following assumptions.

%\begin{assumption}\label{ass_add_sep}
%	The error term $ U $ is additively separable.
%\end{assumption} 

\begin{assumption}[Regularity]\label{ass_regular}
	The functions $h$, $\widetilde{h}$, $f_{X|W}$, $f_{Y|W}$, $f_{\varepsilon}$, and $f_{\widetilde{\varepsilon}}$ are three times differentiable. 
\end{assumption}

\begin{assumption}[Nonlinearity]\label{ass_nonlinear}
	The functions $ h$ and $\tilde{h}$ are nonlinear in their first arguments.
%	 for each $ \w $.
%	 in the support of $\W$.
\end{assumption}

The nonlinearity of regression functions (Assumption \ref{ass_nonlinear}) can be used to make inference on the causal structure of a model is obtained first by \textcite{hoyer09anm}.
We extend their work by allowing for additional control variables $W$ and also considering heteroskedasticity of the error term with respect to these covariates. Intuitively, nonlinearity of $h$ ensures that the error terms in the reverse model are not independent of the regressor, which provides power of the test. While linear models are used in many economic applications, they are typically seen as approximations of nonlinear relationships between dependent variable and regressors.

\subsection{Testability}

%\footnote{Hatten wir keine Diskussion bzgl nonlinear $h$? Zb: Econometric models are often linear as plausible approximations to more complex economic relationships. Assumption \ref{ass_nonlinear} restricts the true underlying relation to be of nonlinear form, which is a mild restriction in most economic environments. In Monte Carlo simulations section, we investigate the effect of nonlinearity on the power to detect the correct causal direction.}

%\footnote{Was koennen wir sagen, wenn das Modell tatsaechlich linear ist?}
%\begin{assumption}[Heteroskedasticity]\label{ass_het}
%	Assume
%	\begin{equation}\label{eq_heterosk}
%	U = \sigma(\W) \; \varepsilon\;  \; \text{with}\;  \; \varepsilon \indept (X,\W)
%	\end{equation}
%	for some strictly positive function $ \sigma(\cdot)$.
%\end{assumption}
%Existing causal discovery algorithms do not take into account heteroskedastic error structures. Assumption \ref{ass_het} explicitly introduces such heteroskedasticity of the error term with respect to the control variables $ \W $. Note that Assumption \ref{ass_het} implies $  U \indept X|\W  $.
We are now in a position to formulate the main theorem of this paper.
	\begin{theorem}\label{thm}
	Let Assumptions \ref{ass_regular} and \ref{ass_nonlinear} be satisfied. Both model \eqref{eq_w_causal} and the reverse model \eqref{eq_w_backw1} exist only if $\xi(x,w) := \log f_{X|W}(x|w)$ satisfies the linear inhomogeneous differential equation
	\begin{align}\label{eq_diff}
\frac{\partial^3\xi(x,\bar{\w})}{\partial x^3}=\frac{\partial^2\xi(x,\bar{\w})}{\partial x^2}G_1(x,\bar{y},\bar{\w})+G_2(x,\bar{y},\bar{\w})
	\end{align}
	for all $x$ and some fixed $(\bar y,\bar w)$, where $ G_1(x,\bar{y},\bar{\w}) $ and $ G_2(x,\bar{y},\bar{\w}) $ are defined in Appendix~\ref{app_proof}. 
\end{theorem}

The proof of this statement can be found in Appendix~\ref{app_proof}. For the proof of the result, we build on \textcite{hoyer09anm} and extend their result to allow for control variables $W$ with additional form of heteroskedasticity. Intuitively, it is shown that causal and anticausal models can only exist simultaneously under very specific circumstances: if the joint distribution of $(Y,X,W)$ satisfies both a causal and a anticausal model, we can show that densities $\log f_{X|W}$ and $\log f_{\varepsilon}$ of the causal model have to satisfy the linear inhomogeneous differential equation~\eqref{eq_diff}. The solutions of this differential equation restrict the log density of $ X $ given $W$ to lie in a (specific) three-dimensional space, although a priori the (generic) space of possible log marginal densities of $ X $ given $W$ is infinite-dimensional \citep[this argument follows][closely]{hoyer09anm}. To achieve testability of reverse causality, we need to exclude those distributions that satisfy the differential equation, i.e., we need to exclude specific combinations of $\log f_{X|W}$, $\log f_{\varepsilon}$, and $h(\cdot)$.

\begin{remark}[Characterization of differential equation \eqref{eq_diff}]
In Table \ref{table:testable}, we reproduce an exhaustive list of all model specifications that satisfy the differential equation \eqref{eq_diff} by \textcite{zhang09pnl} under the additional assumption that the error $\varepsilon$ has large support. The list of $(f_{X},f_{\varepsilon},h(\cdot))$ tuples that satisfy the differential equation is even smaller than in \textcite{zhang09pnl} because we constrain the model space by assuming a nonlinear $h$. %When $W$ is not included in the model,  list the only three
%\pcomment{only three, not five as in \textcite{zhang09pnl}, because we can use our assumption that h is nonlinear here} specific tuples that would satisfy the differential equation under Assumption \ref{ass_nonlinear}. 
%For instance, the case where $ \log f_{X|W}(x) = c_{1} e^{c_{2} x}+c_{3} x+c_{4} $ (where $c_1$, $c_2$, $c_3$, and $c_4$ are constants), $ \log f_{\varepsilon} $ is a one-sided asymptotically exponential distribution, and $h(\cdot)$ is strictly monotonic while its first derivative is going to zero as $x \rightarrow -\infty$ and $x \rightarrow \infty$ is  excluded by the differential equation in Theorem \ref{thm}. Enlisting one example of these specific conditions that the $(f_{X|W},f_{\varepsilon},h(\cdot))$ tuple needs to fulfill to satisfy the differential equation underscores that the tuples we need to exclude are not only limited in number but tuned to each other in a specific way. 
%In other words, testability of the causal direction is achieved as long as the densities for $f_{X|W}$ and $f_{\varepsilon}$ as well as $h(\cdot)$ are not tuned to each other in a specific way to satisfy the differential equation.
It is remarkable that in each specification $I$, $II$, and $III$, the density of $\varepsilon$ is not even integrable. Even more, $E[\varepsilon]$ does not exist. This illustrates that even though solutions to the differential equation \eqref{eq_diff} can be computed, they are not relevant in most (if not all) empirical applications.
\end{remark}

\begin{table}[h!]
 \begin{center}
 \setlength{\tabcolsep}{1pt} % Default value: 6pt
  \renewcommand{\arraystretch}{1.5}
 \begin{tabular}{|c||c|c|}
\hline & $f_\varepsilon$ & $f_X$  \\
\hline
\hline I& $f_\varepsilon(u)=c_{1} \exp(c_{2} u)+c_{3} u+c_{4} $ & $(\log f_X(x))'\to c_1\neq 0$ as $x\to+\infty$ or $x\to-\infty$ \\
\hline II & $f_\varepsilon(u)=c_{1} \exp(c_{2} u)+c_{3} u+c_{4} $ &$f_X(x)=\{c_{1} \exp(c_{2} x)+c_{3} \exp(c_{4} x)\}^{c_5} $  \\
\hline III & $f_\varepsilon(u)=\{c_{1} \exp(c_{2} u)+c_{3} \exp(c_{4} u)\}^{c_5} $ & $\lim\limits_{x\to-\infty}(\log f_X(x))'= c_1\neq 0$, $\lim\limits_{x\to+\infty}(\log f_X(x))'= c_2\neq 0$ \\
\hline
\end{tabular}
\end{center}
\vskip -.4cm
 \caption{{\small All situations in which reverse causality is not testable. Constants $c_1,c_2,c_3,c_4,c_5$ might be different in different cases. In all situations,  $h$ strictly monotonic, and $h^{\prime}(x) \rightarrow 0$, as $x \rightarrow+\infty$ or as $x \rightarrow-\infty$ under the maintained assumption of large support of $\varepsilon$. }}
\label{table:testable}
 \end{table}

% Intuitively, it is shown that causal and anticausal models can only exist simultaneously under specific circumstances: if the joint distribution of $ X $ and $ Y $, $ f_{X,Y}(x,y) $, is to allow for both a causal and an anticausal model, we can show that the causal model has to satisfy the linear inhomogeneous differential equation \eqref{eq_hets_differential_eq}. That $ f_{X,Y}(x,y) $ has to be fine-tuned in this specific manner implies that, in general, causal and anticausal model cannot be satisfied simultaneously.

The next result provides a more concrete formulation of Theorem \ref{thm} and how it can be applied to model specification testing. This corollary follows immediately from Theorem \ref{thm} and its proof is thus omitted. 
\begin{corollary}\label{cor:test}
	Let Assumptions \ref{ass_regular} and \ref{ass_nonlinear} be satisfied.
Then model 	\eqref{eq_w_causal} rules out the reverse model \eqref{eq_w_backw1} if the joint distribution of $(X,W)$ does not satisfy the differential equation \eqref{eq_diff}.
%	 then the true data-generating process satisfying model \eqref{eq_w_causal} rules out the reversed model \eqref{eq_w_backw1}.
%	
\end{corollary}
Corollary  \ref{cor:test} allows identification of the causal direction from observational data by analyzing to what extent the independence of errors and covariates holds. %Therefore, we can infer whether the causal or anticausal model correctly describes the causal structure by testing $ U \indept X|\W $ and $ \widetilde{U} \indept Y|\W $ and rejecting the model whose corresponding independence relation is rejected. 
The nonlinearity of $ h $ and the additive separability of the error term, $ U $, give the proposed test power.

Note that even when $h$ is linear ($Y = X + U$, simplifying by abstracting from $W$), it is only possible to rewrite this as a reverse model, $X = Y + \widetilde{U}$, with $Y$ independent of $\widetilde{U}$, if all variables are Gaussian. If at most one of the exogenous variables is non-Gaussian, residual and purported cause are dependent in the reverse model \citep{shimizu06lingam}.

%We conclude this section with some further observations about the causal discovery literature. Next to the \textit{a priori} restriction of the model class, which we follow in this paper, there are more proposals to identify causal directionality. First, there is work relying on information-geometric arguments: the essential idea is that the conditional distribution of the effect given its cause does not contain information about the marginal distribution of the cause \parencite{janz10algo}. The information content is formalized using the notion of Kolmogorov complexity, which in turn is approximated by the entropy of underlying probability distributions. Second, there are constraint-based causal discovery algorithms. These methods construct a causal model based on an exhaustive list of statistical independencies of any two observed variables conditional on sets of the other observed variables \parencite{peters14}. One needs at least three observed variables to apply such methods. Thus, the bivariate nature of the problem we are addressing precludes the application of constraint-based causal discovery algorithms. Furthermore, there are score-based methods that compare, e.g., penalized likelihoods across models and base inference of causal direction thereon \parencite[see][for an example]{nowzohour16score_based_learning}. 

\subsection{Implementation the Reverse Causality Test}
Algorithm \ref{algorithm_test} shows detailed steps of the implementation of the test. In words, after some pre-processing (Step 1 and 2), we propose estimating a nonlinear model in both directions, i.e. with $ Y $ and $ X $ as dependent variables respectively (Step~3), calculating residuals for both models (Step~4) and testing for conditional independence using a kernel conditional independence test (KCI) introduced by \textcite{zhang11conditional} (Step~5). We discuss this test statistic, see eq. \eqref{eq_kci_pop}, and its development in Section~\ref{sec_ci_test}; see the detailed discussion in Section \ref{sec_discussion}.
\begin{algorithm}[b!]
	\RestyleAlgo{boxed}
	\KwData{$ \mathcal{D} =  \{Y_i,X_i,\W_i\}_{i=1}^n $} 
	\KwIn{hyperparameters for KCI test: heuristic kernel bandwidth $ \lambda $, set at $ \lambda =  0.8 $ if sample size $ n \leq 200 $, $ \lambda =  0.3 $ if sample size $ n > 1200 $ and $ \lambda = 0.5 $ otherwise; heuristic regularization parameter $ \lambda_R $ is set to $  10^{-3}$
%	, nominal size of conditional independence test: $\alpha\in(0,1)$
	}
	\KwOut{Decision whether to reject the model in \eqref{eq_w_causal}} 
	\textbf{Step 1}: Normalize data to have mean equal to zero and variance equal to one. \\
	\textbf{Step 2:} Randomly split data in half to form training $\mathcal{D}^{tr} = \{Y_i,X_i,\W_i\}_{i=1}^{n/2} $ and test set $\mathcal{D}^{te} = \{Y'_i,X'_i,\W'_i\}_{i=(n/2)+1}^n $ \\
	\textbf{Step 3}: Estimate generalized additive models (GAMs) based on $\mathcal{D}^{tr} $ \\
	GAM1: $ Y=h(X,\W) + U $, call resulting estimate $ \widehat{h} $ \\
	GAM2: $ X=\widetilde{h}(Y,\W) + \widetilde{U} $, call resulting estimate $ \widehat{\widetilde{h}} $ \\
	\textbf{Step 4}: calculate residuals based on $\mathcal{D}^{te} $ \\
	$\widehat{U} := Y'-\widehat{h}(X',\W') $, and  \\
	$\widehat{\widetilde{U}} := X'-\widehat{\widetilde{h}}(Y',\W') $ \\
	\textbf{Step 5}: Test conditional independence with KCI test (based on residuals from Step~4) \\
	use $ \widehat{U} $, $ X' $ and $ \W' $ to test $ U \indept X | \W $ with $ \mathrm{KCI} $ test whose null hypothesis is conditional independence; call resulting \textit{p}-value $ p_\text{model} $ \\
	use $ \widehat{\widetilde{U}} $, $ Y' $ and $ \W' $ to test $ \widetilde{U} \indept Y | \W $ with $ \mathrm{KCI} $ test whose null hypothesis is conditional independence; call resulting \textit{p}-value $ p_\text{reverse} $ \\
	\textbf{Step 6}: Decide to reject model \eqref{eq_w_causal} if $p_\text{model} < \alpha $.
	\caption{Reverse causality test at nominal level  $\alpha\in(0,1)$}\label{algorithm_test}
\end{algorithm}

%\textcolor{blue}{What is $p_\text{model} $? The $p$-values? How are they calculated?}\pcomment{yes, $p_\text{model} $ is the p value for the condtl independence test of residuals and purported cause in the \textit{model} (which used to be called 'causal model'), how they are calculated is discussed at length in Sect 2.5 Testing Conditional Independence, see also \textcite{zhang11conditional} [this comment can be deleted from my perspective]}

\subsection{Causal Discovery}

Alternatively, instead of imposing model \eqref{eq_w_causal} as a maintained hypothesis, we can also remain agnostic about the true causal relationship and infer which of the models is the correct one. Formally, this requires assuming that either model \eqref{eq_w_causal} or \eqref{eq_w_backw1} hold. This is formalized in the following corollary which follows immediately from Theorem \ref{thm} and its proof is thus omitted. 

\begin{corollary}\label{cor:ex}
	Let Assumptions \ref{ass_regular} and \ref{ass_nonlinear} be satisfied.
Suppose either the model 	\eqref{eq_w_causal} or the reverse model \eqref{eq_w_backw1} holds. Then, if the joint distribution of $(X,W)$ does not satisfy the differential equation \eqref{eq_diff}, the true model is identified. 
%, and vice versa, if the joint distribution of $(X,W)$ does not satisfy the differential equation \eqref{eq_diff}.
%	 then the true data-generating process satisfying model \eqref{eq_w_causal} rules out the reversed model \eqref{eq_w_backw1}.
%	
\end{corollary}

\begin{algorithm}[h!]
	\RestyleAlgo{boxed}
	\KwData{$ \mathcal{D} =  \{Y_i,X_i,\W_i\}_{i=1}^n $} 
	\KwIn{hyperparameters for KCI test: heuristic kernel bandwidth $ \lambda $, set at $ \lambda =  0.8 $ if sample size $ n \leq 200 $, $ \lambda =  0.3 $ if sample size $ n > 1200 $ and $ \lambda = 0.5 $ otherwise; heuristic regularization parameter $ \lambda_R $ is set to $  10^{-3}$.}
	\KwOut{Decision whether true causal model is $ X \rightarrow Y $ or $ Y \rightarrow X $} 
	\textbf{Step 1}: Normalize data to have mean equal to zero and variance equal to one. \\
	\textbf{Step 2:} Randomly split data in half to form training $\mathcal{D}^{tr} = \{Y_i,X_i,\W_i\}_{i=1}^{n/2} $ and test set $\mathcal{D}^{te} = \{Y'_i,X'_i,\W'_i\}_{i=(n/2)+1}^n $ \\
	\textbf{Step 3}: Estimate generalized additive models (GAMs) based on $\mathcal{D}^{tr} $ \\
	GAM1: $ Y=h(X,\W) + U $, call resulting estimate $ \widehat{h} $ \\
	GAM2: $ X=\widetilde{h}(Y,\W) + \widetilde{U} $, call resulting estimate $ \widehat{\widetilde{h}} $ \\
	\textbf{Step 4}: calculate residuals based on $\mathcal{D}^{te} $ \\
	$\widehat{U} := Y'-\widehat{h}(X',\W') $, and  \\
	$\widehat{\widetilde{U}} := X'-\widehat{\widetilde{h}}(Y',\W') $ \\
	\textbf{Step 5}: Test conditional independence with KCI test (based on residuals from Step~4) \\
	use $ \widehat{U} $, $ X' $ and $ \W' $ to test $ U \indept X | \W $ with $ \mathrm{KCI} $ test; call resulting test statistic $ \mathrm{KCI}_\text{causal} $ \\
	use $ \widehat{\widetilde{U}} $, $ Y' $ and $ \W' $ to test $ \widetilde{U} \indept Y | \W $ with $ \mathrm{KCI} $ test; call resulting test statistic $ \mathrm{KCI}_\text{anticausal} $ \\
	\textbf{Step 6}: Decide on causal direction \\
	\uIf{$ \mathrm{KCI}_\text{causal} < \mathrm{KCI}_\text{anticausal} $}{
		accept $ X \rightarrow Y $ as correct model
	}
	\uElseIf{$ \mathrm{KCI}_\text{causal} >  \mathrm{KCI}_\text{anticausal} $}{
		accept $ Y \rightarrow X $ as correct model
	}
	\ElseIf{$ \mathrm{KCI}_\text{causal} = \mathrm{KCI}_\text{anticausal} $}{
		inconclusive result
	}

	\caption{Bivariate causal discovery with control covariates}\label{algorithm_causal_disc}
\end{algorithm}
Algorithm \ref{algorithm_causal_disc} shows detailed steps of the implementation of the bivariate causal discovery which is motivated by \ref{cor:ex}. Steps 1 to 4 are the same as in \ref{algorithm_test}. In step 5, we compute the test statistics corresponding to two conditional independence tests: one for model \eqref{eq_w_causal} and one for \eqref{eq_w_backw1}. The relative size of the resulting test statistics is informative about which model is the correct causal model (Step~6).

\begin{figure}[h!]
	\centering
	\includegraphics[width=.8\textwidth]{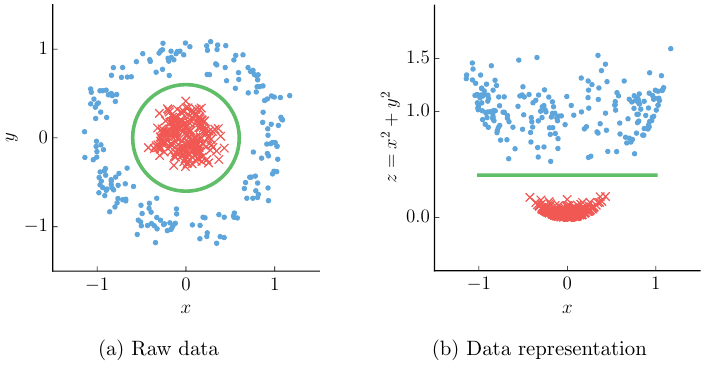}
	\caption[A nonlinear classifier]{\textbf{A nonlinear classifier. }Panel (a): data can only be separated by a nonlinear decision boundary (green circle), a linear algorithm fails. Panel (b) mapping the data to a higher-dimensional space by introducing an additional feature $ z = x^2 + y^2 $ enables a linear decision boundary (green line) to separate the data. Note that the $y$ dimension is ommited in (b); it extends perpendicular to the 2D sheet of paper/screen. In a 3D graph, the data looks like an inverted cone with a rounded bottom. Figure credit: \textcite[][Figure 3.1]{dlp16}}
	\label{fig_classifier}
\end{figure}

\subsection{Testing Conditional Independence}\label{sec_ci_test}

This section introduces the concept of Hilbert Space embeddings of probability distributions and their use for (un)conditional independence testing of random variables. Since this notion is not common in the econometrics literature and conditional independence testing forms a central part of the proposed algorithm, we discuss the procedure in detail. We first intuitively introduce important underlying concepts such as feature maps, reproducing kernel Hilbert spaces, etc. keeping technical details to a minimum before turning to how these constructs can help to formulate a conditional independence test. See Appendix \ref{app_rkhs} for the formal statements.

\subsubsection{Feature maps}\label{feature_maps}
To introduce the usefulness of a feature map, consider the following problem. Terms used loosely in this paragraph are precisely defined below. Imagine you want to distinguish between two groups of subjects each characterized by two dimensions, say $x=$ weight and $y=$ height, by using a linear classifier (i.e. a linear regression that serves as a boundary between the two classes). If the data looks like those in Figure \ref{fig_classifier}(a), a linear classifier will perform poorly since there is no linear decision boundary that it could uncover. A solution to the problem lies in mapping the data from two-dimensional input space to a higher-dimensional feature space by introducing an additional feature $ z = x^2 + y^2 $ that complements existing features $ x $ and $ y $ (here the map is from a two-dimensional to a three-dimensional space; in practice the feature space will have many more dimensions). In this higher-dimensional space, there is a linear boundary that separates the two classes, see Figure \ref{fig_classifier}(b). This example is adopted from \textcite{dlp16}.

%\begin{figure}[t]
%	\centering
%	\includegraphics[width=.8\textwidth]{\revcauspath/manuscripts/graphics/nonlinear_classifier_example.pdf}
%	\caption{Panel (a): data can only be separated by a nonlinear decision boundary, a linear algorithm fails. Panel (b) mapping the data to a higher-dimensional space by introducing an additional feature $ z = x^2 + y^2 $ enables a linear algorithm to separate the data, source: \textcite[][Figure 3.1]{dlp16}}
%	\label{fig_classifier}
%\end{figure}

Similarly to the linear classifier in Figure \ref{fig_classifier}(a) that does not succeed in distinguishing between two classes that are separated by a nonlinear decision boundary in input space, the (linear) covariance between two random variables does not succeed in detecting nonlinear statistical dependencies. Mapping the data from input to feature space enables the exemplary classifier to linearly describe the decision boundary in feature space despite it being nonlinear in input space. Similarly, one can use the theory on reproducing kernel Hilbert spaces (RKHS) to construct a representation of marginal and conditional probability distributions in higher-dimensional feature space. The covariance operator between two random variables in that feature space is then informative about nonlinear dependencies in input space. In sum, \textit{any linear algorithm in high-dimensional feature space corresponds to a nonlinear algorithm in input space}. Crucially, inner products between feature space representations can be estimated without knowing the exact feature representation itself (the so-called `kernel trick'). We now turn to a formal definition of a RKHS and kernel mean embedding of probability distributions.

%\begin{figure}[t]
%	\centering
%	\includegraphics[width=.5\textwidth]{feature_map.pdf}
%	\caption[Illustration of feature map ]{\textbf{Illustration of feature map $ \Phi $.} Each data point $ x $ in input space is mapped to a function $ \phi(x) $ in feature space, which represents $ x $ in terms of its similarity to all other data points. Figure credit: \textcite[][p. 32]{kernels01}}
%	\label{fig_feature_map}
%\end{figure}

\subsubsection{Kernels as inner product of implicit feature map}\label{sec_construction_rkhs}

In practice, instead of manually defining a set of appropriate features (such as $ z = x^2+y^2 $ in the previous example), flexible functions can be used to define the feature map. 
Formally, we define a feature map $ \Phi $ from input space $ \mathcal{X} $ to the space of functions $ \mathbb{R}^\mathcal{X} $:
\begin{align*}
\Phi:\;\; \mathcal{X} &\rightarrow \mathbb{R}^\mathcal{X} \\
x &\mapsto k(\cdot,x)
\end{align*}
where $ k $ is a positive-definite kernel\footnote{A positive-definite kernel is a kernel with an associated kernel matrix $ K $, which has entries $ K_{ij} := k(x_i,x_j) $, that is positive-definite.}, such as the Gaussian kernel, which is defined as
\begin{equation}\label{eq_gaussian_kernel}
	k(v,v') := \exp \Big(-\frac{\norm{v-v'}_{\ell_2}^2}{\lambda}\Big)
	\end{equation}
for arbitrary vectors $ v $ and $ v' $ and a bandwidth parameter $ \lambda>0$, where $ \norm{\cdot}_{\ell_2} $ denotes the $\ell_2$ norm. Each data point can thus be richly represented by its similarity (defined by the kernel) to all other data points. It can be shown that the inner product of two such feature maps in an RKHS reduces to an evaluation of the kernel itself \parencite[see][and Appendix \ref{app_rkhs}]{kernels01}:
\begin{equation*}
\langle k(\cdot,x), k(\cdot,x') \rangle = \langle \Phi(x),\Phi(x')\rangle = k(x,x').
\end{equation*}

This result shows that the inner product of possibly infinite-dimensional feature representations, $ \langle \Phi(x),\Phi(x')\rangle $, can be evaluated through the kernel $ k $ without making the feature representation explicit (the so-called `kernel trick' in machine learning). Any algorithm or other data processing technique that relies on calculating inner products between data representations can be `kernelized,' i.e. transformed into a nonlinear algorithm by mapping the data into a higher-dimensional Reproducing Kernel Hilbert Space. The covariance, which can be defined as a dot product, falls into this category.

Instead of representing a specific data point by means of a feature vector, we subsequently intend to represent a whole probability distribution in terms of a higher-dimensional vector. One way to think about this procedure intuitively is to note that probability distributions can be characterized uniquely by an infinite sequence of their moments. Thus, the elements of the infinite-dimensional feature vector can be populated by moments of increasing order when embedding a probability distribution in the RKHS, which gives rise to a unique representation of the probability distribution.

\subsubsection{Partial cross-covariance operators and conditional independence}

The conditional independendece test we use relies on a characterization of conditional independence as a vanishing partial cross-covariance operator between two RKHSs. To get an intuition, consider an analogy to the characterization of conditional independence for jointly Gaussian variables in terms of vanishing partial correlation. First note that, for jointly Gaussian variables $ (Z_1,Z_2,Z_W) $, the conditional independence, $ Z_1 \indept Z_2 | Z_W $ can be characterized as the correlation between $ Z_1|Z_W $ and $ Z_2|Z_W $ being zero. Partial correlation is a \textit{linear} concept defined by the orthogonality of \textit{linear} maps of $ Z_1 $ and $ Z_2 $ on the space orthogonal to $ Z_W $. It can only characterize conditional independence for jointly Gaussian variables because of the linearity of the underlying maps. Intuitively, one can extend the results to apply to \textit{nonlinear} dependence of \textit{arbitrarily} distributed random variables if such maps can be described more flexibly. We have seen how maps of data into higher-dimensional RHKS enables the use of linear algorithms to study nonlinear relationships. This reasoning also underlies the following characterization of conditional independence for arbitrarily distributed random variables.

\textcite{daudin80} establishes the equivalence
\begin{equation}\label{eq_daudin}
\sup_{f \in \mathcal{F}_{\widetilde{X}},\, g \in \mathcal{F}_{{U}} } \mathbb{E}[{f}(\widetilde{X}){g}({{U}})] = 0 \Leftrightarrow X \indept U | W,
\end{equation} 
for properly chosen function spaces, based on function spaces $
\mathcal{F}_{\widetilde{X}} := \big\{{f} \in L^2_{\widetilde{X}} : \; \mathbb{E}[{f(\widetilde{X})}|W]= 0\big\} $ and $	\mathcal{F}_{\widetilde{U}} := \big\{{g}: \; g(\widetilde{U}) = \check{g}({U}) - \mathbb{E}[\check g(U)|W]\text{ where } \check{g} \in L^2_U \big\} $ where for any random variable $Z$ the Hilbert space $ L^2_Z =\{f:\, \mathbb{E}[f^2(Z)]<\infty\} $.
	
%However, this result cannot be used to derive a feasible procedure to test conditional independence because these function spaces are too large to search over. To make the result feasible,
Throughout the remainder of this section, we consider continuous random variables $ X $, $ U $ and $ W $ with domains $ \mathcal{X} $, $ \mathcal{U} $ and $ \mathcal{W} $, and with positive definite kernels $ k_\mathcal{X} $, $ k_\mathcal{U} $, and $ k_\mathcal{W} $ defined on these domains. These give rise to RKHSs $ \mathcal{H}_\mathcal{X} $, $ \mathcal{H}_\mathcal{U} $ and $ \mathcal{H}_\mathcal{W} $ respectively. Further, we make use of the notation $ \widetilde{X} = (X,W) $ and $ \widetilde{U} = (U,W) $ and define $ k_{\widetilde{\mathcal{X}}} = k_\mathcal{X}\times k_\mathcal{W} $ and corresponding RKHS $ \mathcal{H}_{\widetilde{\mathcal{X}}} $. Denote the feature maps corresponding to RKHS $ \mathcal{H} $ with $ \phi_\mathcal{H} $. To reduce the complexity of Daudin's equivalence result, \textcite{zhang11conditional} show that restricting function spaces of $ {f} $ and $ {g} $ to lie in RKHSs $ \mathcal{H}_{\widetilde{\mathcal X}} $ and $ \mathcal{H}_{\mathcal U} $. This restrictions are sufficient to derive an estimable statistic in terms of the Hilbert Schmidt norm of a partial cross-covariance operator:
\begin{equation*}\label{eq_partial_cross_cov_op}
\Sigma_{\widetilde{X}U | W} := \Sigma_{\widetilde{X}{U}} - \Sigma_{\widetilde{X}W}\Sigma_{WW}^{-1}C_{W{U}}
\end{equation*} where $ \Sigma_{\widetilde{X}U} $ is the covariance operator $ \Sigma_{\widetilde{X}U}: \mathcal{H}_\mathcal{\widetilde{X}} \rightarrow \mathcal{H}_\mathcal{U} $ defined as \begin{equation}
\langle \Sigma_{\widetilde{X}U}  \phi_\mathcal{\widetilde{X}},  \phi_{\mathcal{U}} \rangle_{\mathcal H_{\mathcal U}} 	= \mathrm{Cov}(\phi_\mathcal{\widetilde{X}}(\widetilde{X}),\phi_{U}(U)).
\end{equation}
Following \textcite{zhang11conditional}, a vanishing Hilbert Schmidt norm of the partial cross-covariance operator characterizes conditional independence yielding the equivalence result:
\begin{equation}\label{eq_kci_pop}
\big\|\Sigma_{\widetilde{X}{U} | W}\big\|_{HS} = 0 \Leftrightarrow X \indept U | W
\end{equation}
where the Hilbert Schmidt norm of an operator $ A : \mathcal{H}_\mathcal{\widetilde{X}} \rightarrow \mathcal{H}_\mathcal{U}$  is defined as $ \norm{A}_{HS} =   \sqrt{\sum_{j,l \geq 1} \langle f_j,A e_l\rangle^2_{\mathcal{H}_{\mathcal U}}}$ where $ \{e_j\}_{j\geq 1} $ and  $ \{f_j\}_{j\geq 1} $ are an orthonormal basis in $ \mathcal{H}_{\widetilde{\mathcal X}} $ and $ \mathcal{H}_{\mathcal U}$, respectively.

\subsubsection{The KCI test statistic}

\textcite{zhang11conditional} build on the conditional independence characterization in \eqref{eq_kci_pop} and define the KCI test statistic:
\begin{equation}\label{eq_kci_samp}
\text{KCI} = \frac{1}{n}\mathrm{tr}(\widetilde{K}_{\widetilde{X}|W} \; \widetilde{K}_{U|W})
\end{equation}
where $ \widetilde{K}_{\widetilde{X} |W}$ and $\widetilde{K}_{U|W} $ are centralized kernel matrices defined as follows. The centralized kernel matrix for any variable $ Z $ is given by $ \widetilde{K}_Z = HK_Z H $ where $ K_Z $ is the uncentralized kernel matrix, i.e. a matrix whose $ (i,j)  $ element is given by $ k(x_i,x_j) $ where $ k $ is the Gaussian kernel in eq. \eqref{eq_gaussian_kernel}, and $ H = \mathbf{I}_n-n^{-1} \mathbf{1}_n\mathbf{1}_n^\top $ where $ \mathbf{I}_n $ denotes the identity matrix of size $ n $ and $ \mathbf{1}_n $ a vector of ones of length $ n $. These centralized kernel matrices need to be adjusted to reflect the conditioning on $ W $. This is achieved using kernel ridge regression to derive a matrix $R_W$ 
\begin{align*}
R_W = \mathbf{I}_n - \widetilde{K}_W (\widetilde{K}_W + \lambda_R \mathbf{I}_n)^{-1}
\end{align*}
 where $ \lambda_R $ is a regularization parameter.
Finally, the kernel matrix $ \widetilde{K}_{\widetilde{X}|W}$ can be expressed as 
\begin{align*}
 \widetilde{K}_{\widetilde{X}|W} = R_W  \widetilde{K}_{\widetilde X}  R_W.
\end{align*}
Similarly, the construction for $\widetilde{K}_{U|W}$ is analogous and yields 
\begin{align*}
\widetilde{K}_{U|W} = R_W  \widetilde{K}_{U}  R_W.
\end{align*}

Following \textcite{zhang11conditional}, we normalize the data and choose the hyperparameters heuristically as follows. The bandwidth parameter $ \lambda $ is set at $ \lambda =  0.8 $ if sample size $ n \leq 200 $, $ \lambda =  0.3 $ if sample size $ n > 1200 $, and $ \lambda =  0.5 $ otherwise for the construction of $ \widetilde{K}_{\widetilde X} $ and $ \widetilde{K}_{U} $, and at half that value for $ \widetilde{K}_{W} $. The regularization parameter $ \lambda_R $ is set to $  10^{-3}$. These parameters are deemed appropriate when the dimenstionality of $ W $ is small, say less than three. For higher-dimensional $ W $, \textcite{zhang11conditional} recommend choosing different $ \lambda $ and $ \lambda_R $ for $ \widetilde{K}_{\widetilde X} $ and $ \widetilde{K}_{U} $, which can be achieved using cross-validation. 

\textcite{zhang11conditional} derive the asymptotic distribution and corresponding \textit{p}-values of the KCI test statistic under $ H_0: $~conditional independence and show that it achieves pointwise asymptotic level \parencite[see also][]{strobl19}.
In sum, the idea of feature representations motivates the map of the distributions of $ X $ and $ U $ conditional on $ W $ into higher-dimensional spaces where linear correlations correspond to nonlinear dependencies in original space.

\subsection{On Critical Values}\label{sec_discussion}

The theory implies an independence of errors and the covariate in the causal model and a dependence between the errors and the covariate in the anticausal model. The algorithm involves testing the independence between errors and covariate conditional on $ \W $ in both causal and anticausal model. Ideally, the test would conclude with the following decisions: i) if independence can be rejected at a pre-specified significance level in one model but not in the other, one would conclude that the latter model represents the correct causal relation, ii) if independence is rejected in both models, one would conclude that the relation between $ X $ and $ Y $ is confounded, and iii) if independence cannot be rejected in either model, one would conclude that the test does not have sufficient power to decide on the causal direction. It is not possible to implement such a strategy in practice because the true errors are unobserved and the practitioner has to rely on estimated errors. Specifically, the practitioner does not have a sample of $ U := Y - h(X,\W) $ in eq. \eqref{eq_w_causal} at their disposal and, therefore, must rely on estimated errors $ \widehat{U} := Y - \widehat{h}(X,\W) $, and the respective estimated errors of the model in eq. \eqref{eq_w_backw1}, to investigate which model is correct. That these residuals are estimated and, in particular, that they depend on the estimated $ \widehat{h} $, poses a challenge that we discuss now.

\textcite{mooijetal16,hoyer09anm} propose randomly splitting the available data $\mathcal{D} =  \{Y_i,X_i,\W_i\}_{i=1}^{n} $ in training and test sets, denoted $\mathcal{D}^{tr} =  \{Y_i,X_i,\W_i\}_{i=1}^{n/2} $ and $\mathcal{D}^{te} =  \{Y'_i,X'_i,\W'_i\}_{i=(n/2)+1}^{n} $, respectively. $\mathcal{D}^{tr}$ is used to get an estimate $\widehat{h}$ of the true regression function $ h $. $\mathcal{D}^{te}$ is then used to get estimates $\widehat{\varepsilon}' := Y' - \widehat{h}(X',\W')$ of the true errors $\varepsilon$. An error in the estimated $\widehat{h}$ induces a dependence of $\widehat{\varepsilon}'$ and $X'$ (conditional on $ \W' $) even though $\varepsilon$ and $X$ are truly independent (conditional on $ \W $). Consequently, conventional thresholds for the independence test tend to be too loose and would ideally incorporate the fact that $\widehat{h}$ is estimated. Specifically, for a conventional threshold of, say, $ \alpha^* = 0.05 $ the empirical rejection rate will be larger than $ \alpha^* $ in the causal model even though under $ H_0 $ we have that $ U \indept X | \W $, which should lead to an empirical rejection rate roughly equal to $ \alpha^* $. To achieve an empirical size of $ \alpha^* $, one needs to use a threshold $ \alpha = \alpha^* \times \lambda_{\alpha} $ with $ 0 < \lambda_{\alpha} < 1 $. There are no theoretical results on how to choose $ \lambda_{\alpha} $ to account for the dependence of $\widehat{\varepsilon}'$ and $X'$.\footnote{Simulation studies, which are not replicated here, show that $ \lambda_{\alpha} $ depends on the type of distribution that the true error follows. Since there is no way for a practitioner to get a hold on that error distribution, it is impossible to propose rules of thumb, substantiated by simulation exercises, to indicate the level of $ \lambda_{\alpha} $ as a function of observable or estimable quantities.}
However, \textcite{mooijetal16} show that one can infer the correct directionality under additional assumption that the causal \textit{or} the anticausal model exist. Therefore, the identifiability result in Theorem \ref{thm}, which states that either causal or anticausal model, but not both, can satisfy the independence of the error with the covariate, in combination with the existence assumption imposed in Corollary \ref{cor:ex}, which states that either causal or anticausal model exist, allows us to infer the directionality with Algorithm~\ref{algorithm_causal_disc}.

In particular, under the conditions of Corollary \ref{cor:ex}, one can infer that the model with the lower KCI test statistic (i.e. a larger \textit{p}-value of the conditional independence test) is the correct causal model, thereby circumventing the lack of theoretical guidance about an appropriate threshold. %\parencite{mooijetal16}\footnote{This case is similar in spirit to score-based methods such as \textcite{nowzohour16score_based_learning} in the sense that the two KCI statistics are interpreted as scores.}
Making this assumption comes at a cost; namely, a procedure that relies on comparing two test statistics can never conclude that there is not enough information in the data to decide on the causal direction. In other words, such a procedure will never conclude that there is a lack of power to make a decision.

% However, we show in subsequent Monte Carlo simulations that the procedure almost always picks the correct causal model. Assumption~\ref{ass_existence} is strong; yet, if one is willing to make it, the probability of inferring the wrong causal direction is very low.

%A comment on the sample splitting procedure follows. We compare the sample splitting procedure proposed above to an alternative sample splitting procedure to show that the latter is erroneous. For simplicity, we neglect the conditional nature of the applied tests in this paragraph. In this alternative sample splitting procedure $ \mathcal{D}^{tr} $ is used to estimate both $ h $ and the residuals $ \widehat{\varepsilon} := Y - \widehat{h}(X,\W) $. The independence test is then implemented by using $ \widehat{\varepsilon} $ and $\mathcal{D}^{te} =  \{X'_i,\W'_i\}_{i=(n/2)+1}^n $. Here, $ \widehat{\varepsilon} $ and $ X'_i $ are draws of two independent random variables, \textit{by construction}. Moreover, in this alternative approach paired sample tests cannot be used. However, in order to find evidence that favours either the causal or anticausal model, we need to rely on \textit{paired} samples of the estimated residual and the covariate, in both the causal and anticausal models. If the alternative sample splitting were used, the estimated errors were independent of the covariate in \textit{both} causal and anticausal model \textit{by construction} yielding no insight into the causal direction.

\section{Monte Carlo Simulations}\label{sec_mcs}

We investigate finite sample performance of our heteroskedasticity robust reverse causality test and its use in the causal discovery context with Monte Carlo experiments. The results are based on 500 Monte Carlo replications in each experiment and the sample size is varied with $ n \in \{250,500,1000\}$.

We simulate data for the following model: 
\begin{align*}
\begin{pmatrix}
X_i\\
W_i\\
U_i^*
\end{pmatrix} &\sim 
\mathcal N\left(\begin{pmatrix}
0\\
0\\
0
\end{pmatrix},
\begin{pmatrix}
2 &0 &0\\
0 &1 &0\\
0 &0&1
\end{pmatrix}\right)~
\end{align*}
 and 
\begin{equation}\label{eq_error_draw}
U_i = c_{\rho,q}\,\text{sgn}(U_i^*)\,\Big|(1+\phi(W_i))^{\rho/2} U_i^*\Big|^q,
\end{equation}
where $ \phi $ denotes the standard normal probability density function, $\text{sgn}(\cdot)$ is the sign function, and $q,\rho,  c_{\rho,q}$ are constants which vary in the experiments below.
The dependent variable is generated by
\begin{align*}
Y_i&=\kappa_j(X_i,W_i,\tau) + U_i
\end{align*}
where $ j \in \{1,2\} $ and the functions $\kappa_j$ are given by
\begin{align*}
\kappa_1(x,w,\tau) &= x + \tau  x^2 +w, \\
\kappa_2(x,w,\tau) &= x+ \tau \sin(x+\pi/2) + w + w^2.
%/\max\{W_1^2,\dots,W_n^2\}.
%	\kappa_2(x,\tau) &= x + \tau \, x^2 \label{eq_phi2} \\			%	\kappa_2(x,\tau) &= x + \tau \, x^2 \label{eq_phi2} \\
%	\kappa_2(x,\tau) &= sin(x) + \tau \, x^2 \label{eq_phi4}		%	\kappa_2(x,\tau) &= \sin(x) + \tau \, x^2 \label{eq_phi4}
\end{align*}
Here, $ \tau $ controls the degree of nonlinearity between $ Y $ and $ X $. The linear case corresponds to $ \tau = 0 $. 
The parameter $\rho$ captures degree of the heteroskedasticity w.r.t. the control variable $W$. That is, $\mathrm{Var}(U|W=w)= (1+\phi(w))^\rho$ and hence, $ \rho = 0 $ corresponds to the homoskedastic case.
We simulate data for $ \tau \in \{0, 0.25,0.5,0.75,1\} $ and $ \rho \in \{0,1\} $. We rely on the R package \texttt{CondIndTests} \parencite{CondIndTests} for the implementation of the HSIC independence test.

\begin{figure}[t]
	\centering
	\includegraphics[width=1\textwidth]{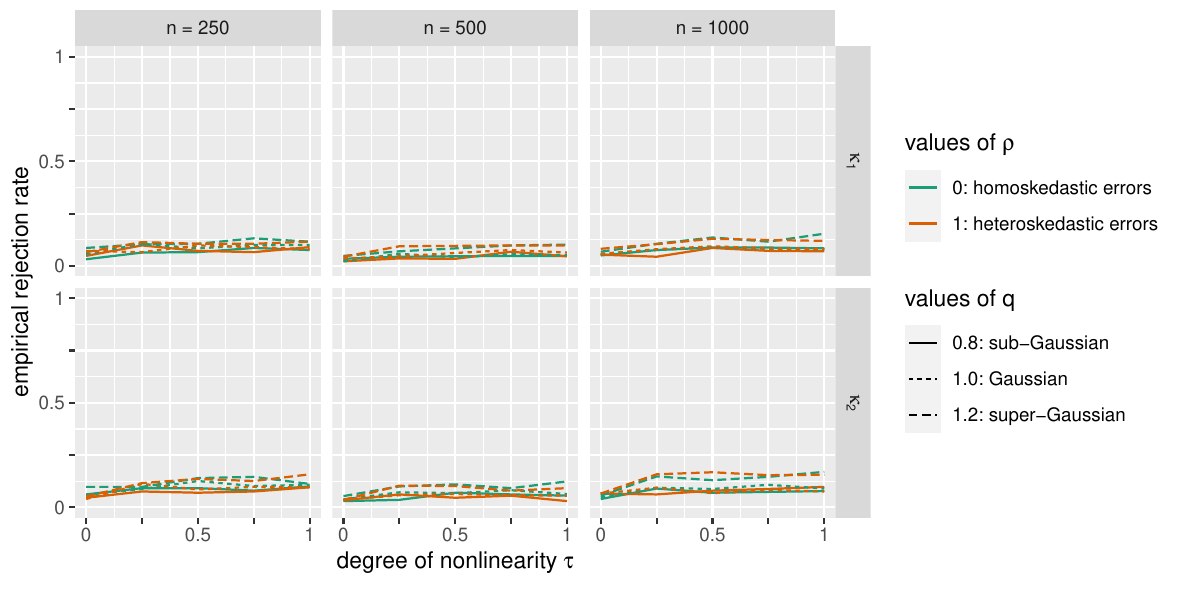}
	\caption{
		Empirical rejection rates of testing conditional independence of covariates and estimated residuals under the model \eqref{eq_w_causal} when $c_{\rho,q}$ is chosen such that $\mathrm{Var}(U) = 1$. 
	}
	\label{fig_mc_var1_causal}
\end{figure}

\begin{figure}[t]
	\centering
	\includegraphics[width=1\textwidth]{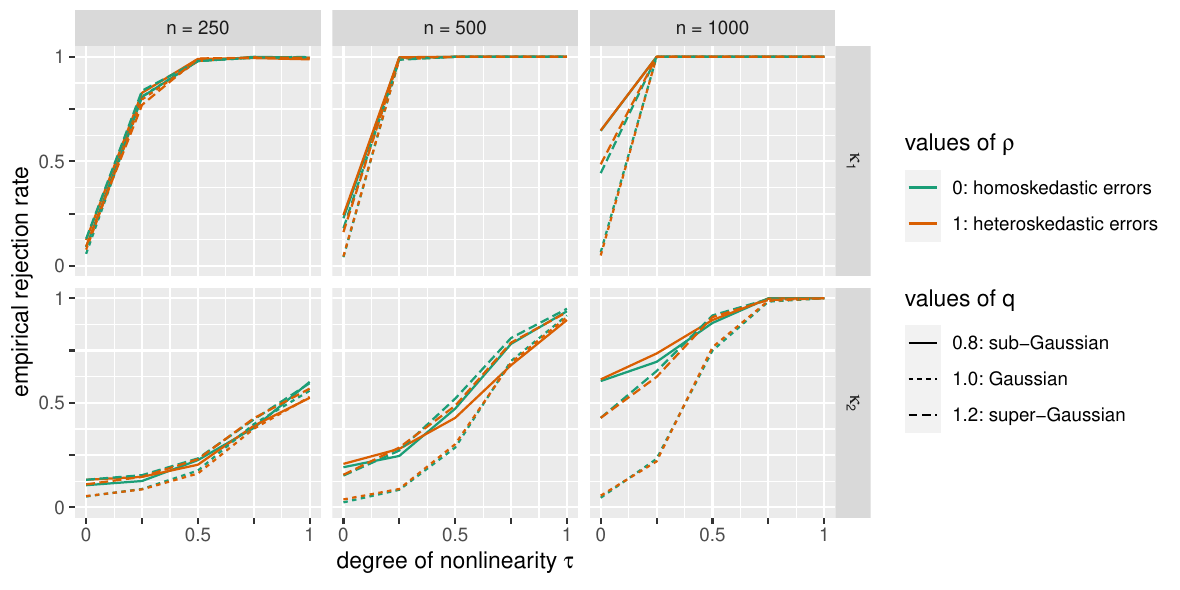}
	\caption{
		Empirical rejection rates of testing conditional independence of covariates and estimated residuals under the reverse model \eqref{eq_w_backw1} when $c_{\rho,q}$ is chosen such that $\mathrm{Var}(U) = 1$. 
	}
	\label{fig_mc_var1_anticausal}
\end{figure}

\begin{figure}[t]
	\centering
	\includegraphics[width=1\textwidth]{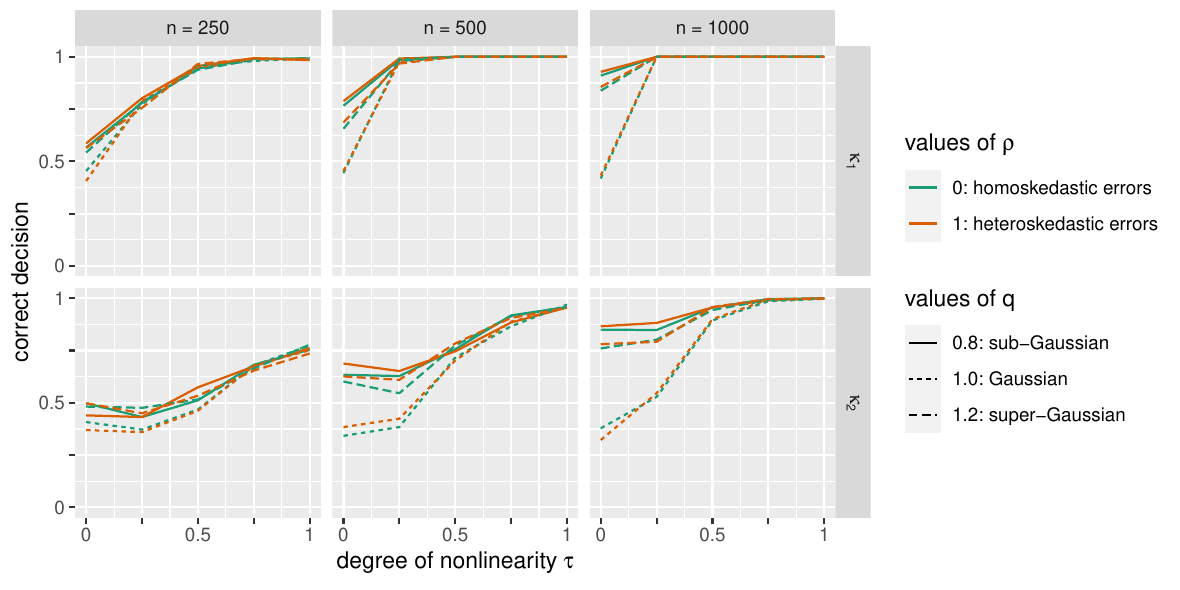}
	\caption{
	Empirical probabilities of correct recovery of the causal direction when $c_{\rho,q}$ is chosen such that $\mathrm{Var}(U) = 1$. 
	 }
	\label{fig_mc_var1_causal_discovery}
\end{figure}

To explore the robustness of our results with respect to the distribution of the error $ {U} $, we run the simulation with errors drawn from sub- and super-Gaussian distributions. We choose the constant $c_{\rho,q}$ (via numerical approximation) such that the variance of $U$ is normalized to one under each choice of $\rho$ and $q$. 
% by raising the draws in \eqref{eq_error_draw} to $ q \in \{0.8,1,1.2\} $ while keeping the sign of each $ U_i $ and the variance of all $ U = \{U_i,\dots, U_n\} $ constant.\ccomment{Nicht ganz klar. Was ist hier gemeint?} $ q=1.2 $ results in super-Gaussian errors. $ q=0.8 $ results in sub-Gaussian errors. 
We estimate both causal and reverse models with a Generalized Additive Model using smoothing splines.
 
\subsection{Testing for Reverse Causality}
We implement the test as described in Algorithm \ref{algorithm_test} for the nominal level $\alpha=0.05$. Figure~\ref{fig_mc_var1_causal} reports the empirical rejection probabilities of testing conditional independence of covariates and estimated residuals under the reverse model \eqref{eq_w_backw1}. Overall, the empirical rejection rates are close to the nominal level for different choices of $\rho$, $q$, and $\tau$. This is remarkable as the testing problem is complex and builds on nonparametric procedures. As such, we cannot expect exact control over the significance level; see also the discussion in Section \ref{sec_discussion}. The test shows some degree of oversizing under super-Gaussian error distributions, indicating the complexity of the testing problem.
 
We illustrate the empirical power of the reverse causality test in Figure \ref{fig_mc_var1_anticausal}. The power of the test, i.e.,  the probability of rejecting the independence of estimated residual and candidate cause in model \eqref{eq_w_backw1}, increases sharply as the degree of nonlinearity $\tau$ increases.
While our identifiability results rely on a nonlinear $h$, see Assumption \ref{ass_nonlinear}, it can be seen that non-Gaussianity can be a source of power similar to nonlinearity: with sufficiently many samples (rightmost column), the empirical rejection rates lie above 0.4 even in the linear case ($\tau=0$) as long as $q\neq1$. This is a well-known result in the causal discovery community, see e.g. \cite{shimizu06lingam}.

%We show the size (Figure \ref{fig_mc_var1_causal}) and power (Figure \ref{fig_mc_var1_anticausal}) for the reverse causality test, see Algorithm \ref{algorithm_test}. The empirical rejection rates for model \eqref{eq_w_causal}, remain low regardless of $\rho$, $q$, and $\tau$. While we do not have exact size control, we show that the empirical size of the test is reasonable.\pcomment{Christoph, kannst Du das bitte formal ueberzeugend schreiben, a la in nonparametric tests is exact size control auch nicht zu erwarten}. See also the discussion on critical values in Section \ref{sec_discussion}. 

%\input{\revcauspath/R_code/mcs_sim/mcs_revcaus_200213_gamma0_unbias1_lambda0_var10fn_form1q_fix20200116.tex}
%\input{mcs_revcaus_200213_gamma0_unbias1_lambda0_var10fn_form1q_fix20200116.tex}

\subsection{Causal Discovery}
In Figure \ref{fig_mc_var1_causal_discovery} we report empirical probabilities of correct classification of the causal direction.
%We show results of the Monte Carlo simulation study in Figure \ref{fig_mc_var1}. 
When the relationship between $ X $ and $ Y $ is linear, i.e. $ \tau = 0 $, and the error Gaussian, i.e. $ q=1 $, the algorithm performs at about chance level. This is consistent with the theory since the causal direction is not identifiable in the linear case.\footnote{Note that \textcite{shimizu06lingam} show that the direction \textit{is} identifiable in the linear case with a non-Gaussian, homoskedastic error distribution, i.e. when $ q \neq 1 $ and $ \rho = 0 $. Though our simulation results suggest that the result also holds with heteroskedastic errors w.r.t. $ W $, we do not extended it formally.} As soon as the relation between cause and effect becomes nonlinear, i.e. $ \tau \neq 0 $, the accuracy of the reverse causal discovery algorithm increases. For instance, when $ n = 500 $ and the relation between cause and $ \tau = 1 $ the algorithm arrives at the correct conclusion in more than 95\% of the Monte Carlo runs, regardless of the level of heteroskedasticity (parameterized by $ \rho $) and shape of the error distribution (parameterized by $ q $). We also see that our procedure has power to detect the correct causal directions in cases where the nonlinearity is less pronounced, i.e., $0<\tau<1$.  Moreover, the performance of the algorithm is robust to changes in the specification of the functional form, which can be seen by comparing the $ \kappa_1 $ and $ \kappa_2 $ rows in Figure~\ref{fig_mc_var1_causal_discovery}. 
%This underscores the robustness to various data-generating processes.
 For a given $ \tau $, the test has more power for $\kappa_1$ because the nonlinear relation between $ X $ and $ Y $ is more pronounced than for $ \kappa_2 $. 
Overall we see that the accuracy of causal discovery increases with sample size, where this change is stronger when the regression function is given by $\kappa_2$.
 We show that the results remain robust to different error variances in Appendix \ref{app_more_results}.Specifically, we consider a low noise regime where $c_{\rho,q}$ is chosen such that $\mathrm{Var}(U)=0.8$ and a high noise regime where $\mathrm{Var}(U)=1.2$. 

In sum, two observations are worth stressing. First, the results show that the algorithm has surprisingly high power when cause and effect are related nonlinearly. Second, the performance of the algorithm does not suffer from heteroskedastic  errors w.r.t. $ W $.

\section{Empirical Illustration}\label{sec_asymmetry}
We use data from the 2013 Survey of Income and Expenditure (``Einkommens- und Verbrauchsstichprobe'', EVS), which is a voluntary survey of roughly 60,000 households in Germany, to test the proposed algorithm. We consider the following variables: income, expenditure, highest educational attainment of the main earner, highest professional training of the main earner, and age group of the main earner. We analyze the causal direction between income and work experience, which we proxy by age group.

Hump-shaped income profiles over the life-cycle are well-documented in labor economics \parencite{heckman06earnings}. It is interesting to test the algorithm for a cause-effect pair where the causal direction is \textit{a priori} clear. Since work experience mechanically increases over the life-cycle, it can be credibly assumed not to be caused by income changes. Therefore, we analyze the directionality between income (Inc) and age where age can be interpreted as proxy for work experience (Exp). We posit the correct causal model to be
\begin{equation}\label{eq_evs_c}
\text{Inc} = h(\text{Exp},W)+\varepsilon_y,
\end{equation} where experience Exp causes income Inc. Vice versa, the anticausal model in which income causes work experience is given as \begin{equation}\label{eq_evs_ac}
\text{Exp} = \widetilde{h}(\text{Inc},W)+\varepsilon_e
\end{equation}
where in each model $ W $ contains all remaining covariates as control (expenditure, highest educational attainment of the main earner, highest professional training of the main earner).

\begin{figure}[H]
	\centering
	\includegraphics[width=.65\textwidth]{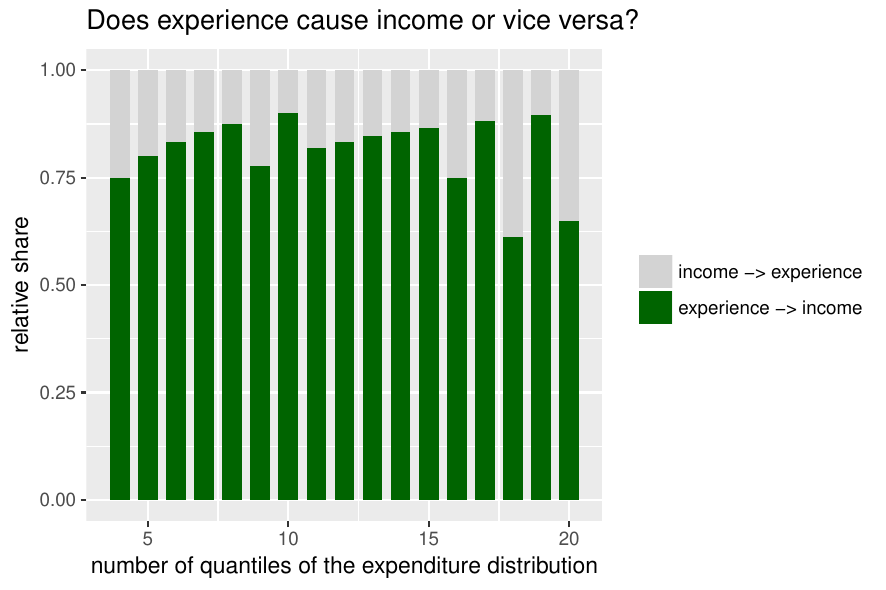}
	\caption[Reverse causality test applied to income $ \leftrightarrow $ work experience, quantiles by expenditure]{This Figure shows the results of the empirical application of Algorithm \ref{algorithm} to the question whether work experience causes income or vice versa. The x-axis shows the number of quantiles that the expenditure distribution is split in (not to be mistaken with the quantiles as such). The stacked bars show the shares of the respective number of quantiles the algorithm decides the causal or anticausal model is the correct model.}
	\label{fig_evs_inc_age_q_exp}
\end{figure}

We aim to alleviate the problem that we are likely to omit many crucial confounding variables by splitting the data in $ n_q $ quantiles of the expenditure distribution. At least part of the omitted confounding factors can be assumed to be fixed within given quantiles as they collect individuals with roughly similar life-styles. This argument applies even more strongly as the expenditure distribution is split into a larger number of quantiles. On the other hand, the larger $ n_q $ the smaller the number of observations within each quantile and the lower the power of the test to identify the correct causal direction. Therefore, we show results for a set of $ n_q = \{4, \dots, 20\} $ quantiles.\footnote{The KCI test, which forms an important part of the algorithm, requires the inversion of $ n \times n $ matrices where $ n $ is the number of observations. Constraints on local computing power preclude running the test on the whole sample with roughly 60,000 observations or with $ n_q = \{1, 2, 3\} $.} For each number of quantiles $ n_q $, we run the causal discovery algorithm in each of these $ n_q $ quantiles and plot the share of quantiles in which the algorithm prefers either model (note that the $ x $-axis in Figure \ref{fig_evs_inc_age_q_exp} refers to the \textit{number} of quantiles the income distribution is split in, not the quantiles as such). For example, the bar above $ n_q = 5 $ in Figure \ref{fig_evs_inc_age_q_exp} denotes that in 4 of the 5 quantiles, i.e. 80\%, the algorithm concludes that experience causes income. Regardless of $ n_q $, the test always favours the model where work experience causes income in at least 50\% of quantiles. The algorithm favours the causal model in more than 75\% of the quantiles for most $ n_q $.
%\begin{figure}[H]
%	\centering
%	\includegraphics[width=.65\textwidth]{evs_200114_age_incqvarinc_qvarinc.pdf}
%	\caption[Reverse causality test applied to income $ \leftrightarrow $ work experience, quantiles by income]{This Figure shows the results of the empirical application of Algorithm \ref{algorithm} to the question whether work experience causes income or vice versa. The x-axis shows the number of quantiles that the income distribution is split in (not to be mistaken with the quantiles as such). The stacked bars show the shares of the respective number of quantiles the algorithm decides the causal or anticausal model is the correct model.}
%	\label{fig_evs_inc_age_q_inc}
%\end{figure}

%Since income is one of the potential causes in this application, splitting the distributions into quantiles of income has the drawback that the variation in income that the algorithm can use within each quantile is mechanically reduced as $ n_q $ increases. Therefore, we replicate the analysis by splitting the data into quantiles of the expenditure distribution. The results can be seen in Figure \ref{fig_evs_inc_age_q_exp}. 
In sum, this application documents that our algorithm gives economically meaningful results in empirical applications.

%\subsection{Discussion}
%
%In both applications we split the income distribution into quantiles and run the test in each quantile. It is conceivable to split the expenditure or age distribution, respectively, in quantiles and run the test in these quantiles. The fact that in the income-expenditure distribution we find suggestive evidence for income being the cause, while in the income-work experience example we find evidence for work experience being the cause, precludes the criticism that our method mechanically concludes that the variable we used to estimate quantiles is the cause.\footnote{pb: dies ist ein wichtiger punkt, weil eine frage die kommen kann/wird, ist warum wir nicht nach expenditure bzw. age quantilen einteilen -- faellt dir noch ein gutes weiteres argument ein, warum wir NUR income quantile anschauen und NICHT expenditure quantile?}

\section{Conclusion}

Endogeneity is a common threat to causal identification in econometric models. Reverse causality is one source of such endogeneity. We build on work by \textcite{hoyer09anm,mooijetal16} who have shown that the causal direction between two variables $ X $ and $ Y $ is identifiable in models with additively separable error terms and nonlinear function forms. We extend their results by allowing for additional control covariates $ \W $ and heteroskedasticity w.r.t. them and, thus, provide a heteroskedasticity-robust method to test for reverse causality. In addition, we show how this test can be extended to a bivariate causal discovery algorithm by comparing the test statistics of residual and purported cause of two candidate models. We extend known results on causal identification and causal discovery to settings with heteroskedasticity with respect to additional control covariates.

An empirical application underscores the feasibility of the proposed algorithm. We analyze the causal link between income and work experience, as proxied by age, and show that our procedure provides evidence that the true causal direction is from work experience to income. Though this result is not substantively surprising because income cannot causally influence work experience, it is encouraging that our algorithm can distinguish between the causal directions without resorting to instruments or other sources of exogenous variation. This underscores the value of our proposed methodology to shed light on the causal structure of economic phenomena without resorting to exogenous variation.

%A limitation that must be addressed in future research is how to adjust the critical values of the involved conditional independence tests, i.e. how to choose $ \lambda_{\alpha} $. Achieving progress in this direction would make Assumption \ref{ass_existence} unnecessary and the provided test even more useful.

%one important point to mention: The usefulness of the proposed algorithm to determine the causal direction between two variables can be illustrated by considering broader structure learning algorithms that can uncover Markov equivalence classes of causal networks. The models that are found in a given Markov equivalence class are characterized by some edges between variables being undirected. A algorithm such as the one proposed here, can be used to direct such undirected edges.

%\newpage

\section{Bibliography}\label{sec_bibliography}
\printbibliography[heading=none]
%
%\pagebreak

%\newpage

\appendix
%\section*{Appendix}

\section{Irreversibility Proof}\label{app_proof}
\begin{proof}[\textbf{\textsc{Proof of Theorem \ref{thm}.}}]
For the proof of the result, we proceed in three steps. First, for the reverse model \eqref{eq_w_backw1}, we establish restrictions on the third derivative of the conditional density of the model. Second, we perform this calculation explicitly for the model \eqref{eq_w_causal}. Third, from the previous steps, we derive distributional restrictions when both models are satisfied.

\medskip
\noindent
\textbf{Step 1.} For the reverse model \eqref{eq_w_backw1}, we first derive an expression for the conditional density of $ X$ given $(Y,W)$, and calculate a third-order derivative of that expression. 
%The resulting derivative is a condition that the anticausal model must fulfill.
Under the reverse model \eqref{eq_w_backw1}, the cumulative distribution function of $ X$ given $(Y,W)$ satisfies
\begin{equation}
\begin{split}\label{eq_density_anticausal}
P(X \leq x|Y = y,\W=\w)&= P(\widetilde{h}(Y,\W)+\widetilde{\varepsilon}\, \widetilde{\sigma}(\W) \leq x | Y=y,\W=\w) \\ 
&=P\Bigg(\widetilde{\varepsilon} \leq \frac{x- \widetilde{h}(Y,\W)}{\widetilde{\sigma}(\W)} | Y=y,\W=\w\Bigg) \\
&=P\Bigg(\widetilde{\varepsilon} \leq \frac{x- \widetilde{h}(y,\w)}{\widetilde{\sigma}(\w)} \Bigg)
\end{split}
\end{equation}
where the last step uses the independence assumption $ \widetilde{\varepsilon} \indept (Y,\W) $.
Thus, we conclude for the conditional probability density functions that
\begin{equation*}
\begin{split}\label{d}
f_{X|Y,\W}(x|y,\w)=f_{\widetilde{\varepsilon}}\Bigg( \frac{x- \widetilde{h}(y,\w)}{\widetilde{\sigma}(\w)} \Bigg)
\end{split}
\end{equation*}
%Under the reversed model \eqref{eq_w_backw1}, the joint density of $X$ and $Y$ conditional on $W$ can be written as 
and, in particular, 
\begin{equation*} \label{eq_het_backw}
f_{X,Y|\W}(x,y|\w)=f_{\widetilde{\varepsilon}}\Bigg( \frac{x- \widetilde{h}(y,\w)}{\widetilde{\sigma}(\w)} \Bigg)f_{Y|\W}(y|\w).
\end{equation*}
We define $\widetilde{\nu}:=\log f_{\widetilde{\varepsilon}}$, $\eta:=\log f_{Y|\W}$ and
\begin{align*}
%\begin{split}
\pi(x,y,\w):&=\log f(x,y|\w)=\eta(y,\w)+ \widetilde{\nu}\Bigg( \frac{x- \widetilde{h}(y,\w)}{\widetilde{\sigma}(\w)} \Bigg)
%\end{split}
\end{align*}
for all $(x,y,w)$ such that $f_{\widetilde{\varepsilon}}\big((x-\widetilde{h}(y,\w))/\widetilde{\sigma}(\w)\big)>0$ and 
$f_{Y|\W}(y,|w)>0$.
Taking partial derivatives yields
\begin{align*}\label{eq_back_der_xy}
\begin{split}
\frac{\partial^2\pi(x,y,\w)}{\partial x \partial y}=- \widetilde{\nu}''\Bigg( \frac{x- \widetilde{h}(y,\w)}{\widetilde{\sigma}(\w)} \Bigg)\frac{\partial \widetilde h(y,w)}{\partial y} \frac{ 1}{\widetilde{\sigma}^2(\w)}
\end{split}
\end{align*}
and
\begin{equation*}\label{eq_backw_der_xx}
\frac{\partial^2\pi(x,y,\w)}{\partial x^2}=\widetilde{\nu}''\Bigg( \frac{x- \widetilde{h}(y,\w)}{\widetilde{\sigma}(\w)} \Bigg)\frac{1}{\widetilde{\sigma}^2(\w)},
\end{equation*}
which, in turn, results in
\begin{equation}\label{eq_ratio_w}
\frac{\frac{\partial^2\pi(x,y,\w)}{\partial x^2}}{\frac{\partial^2\pi(x,y,\w)}{\partial x \partial y}}  = -\frac{1}{\frac{\partial \widetilde h(y,w)}{\partial y}}.
\end{equation}
%\begin{equation}\label{eq_ratio_w}
%\frac{\frac{\partial^2\pi}{\partial x^2}}{\frac{\partial^2\pi}{\partial x \partial y}}= - \frac{\widetilde{\nu}''\Big( \frac{x- \widetilde{h}(y,w)}{\widetilde{\sigma}(w)} \Big)}
%{  \widetilde{\nu}''\Big( \frac{x- \widetilde{h}(y,w)}{\widetilde{\sigma}(w)} \Big) \widetilde{h}'(y,w)}  = -\frac{1}{\widetilde{h}'(y,w)}
%\end{equation}
Therefore, taking the derivative of the ratio in eq. \eqref{eq_ratio_w} w.r.t. $ x $, we conclude
\begin{equation} \label{eq_backw_dx_w}
\frac{\partial}{\partial x} \Bigg( \frac{\frac{\partial^2\pi}{\partial x^2}}{\frac{\partial^2\pi}{\partial x \partial y}} \Bigg)=0.
\end{equation}

\noindent
\textbf{Step 2.} We derive restrictions similar to eq. \eqref{eq_backw_dx_w} for the causal model in eq. \eqref{eq_w_causal}. 
First, we derive an expression for the conditional density of $ Y|X,\W $. Similar to \eqref{eq_density_anticausal}, we can write
\begin{equation*}
\begin{split}
P(Y\leq y|X=x,\W=\w)&= P(h(X,W)+\varepsilon \sigma(\W) < y | X=x,\W=\w) =P\Bigg(\varepsilon \leq \frac{y- h(x,\w)}{\sigma(\w)} \Bigg)
\end{split}
\end{equation*}
which uses the independence assumption $ \varepsilon \indept (X,\W) $.
This lets us conclude for the probability density functions
\begin{equation*}
\begin{split}\label{none}
f_{Y|X,\W}(y|x,\w)=f_{\varepsilon}\Bigg( \frac{y- h(x,\w)}{\sigma(\w)} \Bigg).
\end{split}
\end{equation*}
Therefore, the conditional density of $ X,Y|\W $ can be expressed as
\begin{equation*}\label{eq_het_forw}
f_{X,Y|\W}(x,y|\w)=f_{\varepsilon}\Bigg( \frac{y- h(x,\w)}{\sigma(\w)} \Bigg)f_{X|\W}(x|\w).
\end{equation*}
%with $f_{X|\W}$ and $f_{\varepsilon}$ probability densities on $\mathbb{R}$.
We define $ \nu:=\log \; f_{\varepsilon} $, $ \xi:=\log \; f_{X|\W} $ and
\begin{align*}
\begin{split}
\pi(x,y,\w):&=\log f_{X,Y|\W}(x,y|\w) =\xi(x,\w)+ \nu \Big( \frac{y- h(x,\w)}{\sigma(\w)} \Big)
\end{split}
\end{align*}
for all $(x,y,w)$ such that $f_{\varepsilon}\big((x-h(y,\w))\sigma(\w)\big)>0$ and 
$f_{Y|\W}(y,|w)>0$.
Taking partial derivatives, we conclude
\begin{align}\label{eq_forw_der_xx}
\begin{split}
\frac{\partial^2\pi(x,y,\w)}{\partial x^2}&=  \frac{1}{\sigma^2(\w)}
\left\{ \Bigg(\frac{\partial h(x,w)}{\partial x}\Bigg)^2 \nu''\Bigg(\frac{y-h(x,\w)}{\sigma(\w)}\Bigg) - \frac{\partial^2 h(x,w)}{\partial x^2} \nu'\Bigg(\frac{y-h(x,\w)}{\sigma(\w)}\Bigg) \right\}\\
&\quad+\frac{\partial^2\xi(x,w)}{\partial x^2} \\
&=: \phi_1(x,y,\w) +\frac{\partial^2\xi(x,w)}{\partial x^2}
\end{split}
\end{align}
and
\begin{align}\label{eq_forw_der_xy}
\begin{split}
\frac{\partial^2\pi(x,y,\w)}{\partial x \partial y}&=-  \nu''\Bigg(\frac{y-h(x,\w)}{\sigma(\w)}\Bigg) 
\frac{\partial h(x,w)}{\partial x}
\frac{1}{\sigma(\w)^2} =:\phi_2(x,y,\w).
\end{split}
\end{align}
%In the following derivations we omit the arguments $(y-h(x,w))/\sigma(w)  $ for $ \nu $, $ w $ for $ \sigma $ and $ (x,w) $ for $ \xi $, $ h $. 
In the following derivations we omit the arguments $ (x,\w) $ for $ \xi $, and $ (x,y,\w) $ for $ \phi_1 $ and $ \phi_2 $. 
The ratio of eqs. \eqref{eq_forw_der_xx} and \eqref{eq_forw_der_xy} is given by
\begin{equation}\label{key}
\frac{\frac{\partial^2\pi}{\partial x^2}}{\frac{\partial^2\pi}{\partial x \partial y}} =
\frac{ \phi_1(x,y,\w) +\partial^2\xi(x,w)/\partial x^2}
{\phi_2(x,y,w)}
\end{equation}
which we derive w.r.t. $ x $ to conclude
\begin{align}\label{eq_het_forw_dx}
\begin{split}
\frac{\partial}{\partial x} \Bigg( \frac{\frac{\partial^2\pi}{\partial x^2}}{\frac{\partial^2\pi}{\partial x \partial y}} \Bigg)
&=\frac{\partial^3\xi/\partial x^3}{\phi_2}-\frac{(\partial^2\xi/\partial x^2)(\partial\phi_2/\partial x)}{\phi_2^2}+\frac{(\partial\phi_1/\partial x)\phi_2-(\partial\phi_2/\partial x)\phi_1}{\phi_2^2}.
\end{split}
\end{align}
%\begin{align}\label{eq_het_forw_dx}
%\begin{split}
%\frac{\partial}{\partial x} \Bigg( \frac{\frac{\partial^2\pi}{\partial x^2}}{\frac{\partial^2\pi}{\partial x \partial y}} \Bigg)=&-2h''+\frac{h'^2\nu'''}{\nu''\sigma}+\frac{h'''\nu'\sigma}{\nu''h'}-\frac{\xi'''\sigma^2}{\nu''h'}+\frac{h'^2\nu'''g}{\nu''}\\&-\frac{h''\nu'\nu'''}{\nu''^2}-\frac{h''^2\nu'\sigma}{\nu''h'^2}-\xi''\frac{h'^2\nu'''\sigma+h''\nu''\sigma^2}{\nu''^2h'^2}.
%\end{split}
%\end{align}

%\begin{align}\label{eq_anm8.1}
%\begin{split}
%########
%\end{split}
%\end{align}
\noindent
\textbf{Step 3.} If the reverse model \eqref{eq_w_backw1} holds, we know from \eqref{eq_backw_dx_w} that \eqref{eq_het_forw_dx} must equal zero. By setting \eqref{eq_het_forw_dx} equal to zero and given $h$, $ \nu $, we obtain for each fixed $y$ and $ \w $, which we denote $ \bar{y} $ and $ \bar{\w} $, respectively, a linear inhomogenous differential equation for $\xi$: %(Eq. (4) in \textcite{hoyer09anm})
\begin{equation}\label{eq_hets_differential_eq}
\frac{\partial^3\xi(x,\bar{\w})}{\partial x^3}=\frac{\partial^2\xi(x,\bar{\w})}{\partial x^2}G_1(x,\bar{y},\bar{\w})+G_2(x,\bar{y},\bar{\w}).
\end{equation}
where
$ G_1 = \frac{\partial\phi_2/\partial x}{\phi_2} $ and 
$ G_2 = \frac{(\partial\phi_2/\partial x)\phi_1-(\partial\phi_1/\partial x)\phi_2}{\phi_2} $.
%\begin{equation}\label{eq_anm4}
%	\xi'''=\xi''  \underbrace{ \frac{\phi_2'}{\phi_2} }_\text{$G(x,y,w$)}+\underbrace{\frac{\phi_1\phi_2'-\phi_1'\phi_2}{\phi_2}}_\text{$H(x,y,w)$}
%\end{equation}
%\begin{equation}\label{eq_anm4}
%\xi'''=\xi''  \underbrace{\Big( -\frac{h'\nu'''}{\nu''\sigma}-\frac{h''}{h'} \Big)}_\text{$G(x,y,w$)}+\underbrace{ \frac{2h'^3 \nu'''}{\sigma} + \frac{h'''\nu'}{\sigma} -\frac{\sigma'h''\nu'\nu'''}{\nu''\sigma^2}-\frac{2h'h''\nu''}{\sigma^{2}} -\frac{h''^2\nu'}{h'\sigma} }_\text{$H(x,y,w)$}
%\end{equation}
Making use of the notation  $\chi(x,w):=\partial^2\xi(x,w)/\partial x^2$, we may write
 \begin{equation*}
\frac{\partial \chi (x,\bar{\w})}{\partial x}=\chi(x,\bar{\w})G_1(x,\bar{y},\bar{\w})+G_2(x,\bar{y},\bar{\w}),
\end{equation*} 
which completes the proof.
\end{proof}
Finally, given such a solution for $ \chi(x,\bar{\w}) $ exists, it is given by
\begin{equation*}
\chi(x,\bar{\w})=\chi(x_0,\bar{\w})e^{\int_{x_0}^{x}G_1(\widetilde{x},\bar{y},\bar{\w})d\widetilde{x}}+\int_{x_0}^{x}e^{\int_{\widehat{x}}^{x}G_1(\widetilde{x},\bar{y},\bar{\w})d\widetilde{x}}G_2(\widehat{x},\bar{y},\bar{\w})d\widehat{x}.
\end{equation*}

Following \textcite{hoyer09anm}, we can see that the set of all functions satisfying the condition in eq.~\eqref{eq_hets_differential_eq} is a 3-dimensional affine space. We can fix $\xi(x_0,w)$, $\xi'(x_0,w)$, $\xi''(x_0,w)$ for some arbitrary $x_0$, and $w$ at $\bar{w}$, which determines $\xi$. Given fixed $f$ and $\nu$, and arbitrary $\bar{w}$, the set of all $\xi$ admitting an anticausal model is, thus, contained in a three-dimensional subspace, and therefore not generic. %Loosely speaking, the statement that the differential equation for $\xi$ [with $\xi:=log(p_x)$] has a 3-dimensional space of solutions (while a priori, the space of possible log-marginals $\xi$ is infinite dimensional) amounts to saying that in the general case, our causal model cannot be inverted."
\section{Construction of the RKHS}\label{app_rkhs}

\begin{enumerate}
	\item Generalizing the example illustrated in Figure \ref{fig_classifier}, we consider higher-dimensional feature representations formalized as kernel functions. For instance, consider the Gaussian kernel defined as \begin{equation*}
	k(v,v') := \exp \Big(-\frac{\norm{v-v'}_{\ell_2}^2}{\lambda}\Big).
	\end{equation*}
	 for arbitrary vectors $ v $ and $ v' $ and parameter $ \lambda>0$. This kernel can serve as a higher-dimensional feature representation. In particular, each data point $ x $ is mapped from input space to higher-dimensional feature space where it is represented by its distance to all other data points, i.e. $ k(\cdot,x) $. Each data point is richly represented by its similarity (defined by the kernel) to all other data points.
	
	Formally, we define a feature map $ \Phi $ from input space $ \mathcal{X} $ to the space of functions $ \mathbb{R}^\mathcal{X} $:
	\begin{align*}
	\Phi:\;\; \mathcal{X} &\rightarrow \mathbb{R}^\mathcal{X} \\
	x &\mapsto k(\cdot,x)
	\end{align*}
	where $ k $ is a positive-definite kernel. A positive-definite kernel is a kernel whose associated kernel matrix $ K $, which has entries $ K_{ij} := k(x_i,x_j) $, is positive-definite. Thus, each data point $ x $ is represented by a theoretically infinite-dimensional vector or, in other words, \textit{a function} $ k(\cdot,x) $. In practice, a data point $ x $ is represented by an $ n $-dimensional vector where $ n $ is the number of data points in the sample. %We stress that $ k(\cdot,x) $ can be infinite-dimensional in theory since this flexibility subsequently allows the unique representation of a probability distribution in terms of its infinite moments.
	
	\item The next step in constructing an RKHS is opening the vector space. Consider linear combinations of the feature representations of the form
	\begin{equation*}\label{eq_lin_comb}
	f(\cdot) = \sum_{j = 1}^{m} \alpha_j k(\cdot,x_j)
	\end{equation*}
	for $ \alpha_j \in \mathbb{R} $ and samples $ x_1, \dots, x_m $ of input space $ \mathcal{X} $ where $ m $ is an integer index.
	\item Given a similarly constructed function
	\begin{equation*}
	g(\cdot) = \sum_{l = 1}^{m'} \beta_l k(\cdot,x'_l)
	\end{equation*} with $ \beta_l \in \mathbb{R} $ and samples $ x'_1, \dots, x'_{m'}$ of input space $ \mathcal{X} $ where $ m' $ is an integer index, we can define an inner product between $ f $ and $ g $ as
	\begin{equation}\label{eq_dot_product}
	\langle f,g \rangle := \sum_{j=1}^{m} \sum_{l = 1}^{m'} \alpha_j \beta_l k(x_j,x'_l)
	\end{equation}
	\item  Then complete the space spanned by \eqref{eq_dot_product} by adding the limit points of sequences in the norm defined by $ ||f||:=\sqrt{\langle f,f \rangle} $, the resulting space $ \mathcal{H} $ is called reproducing kernel Hilbert space (RKHS).%\footnote{\parencite[][p. 35]{kernels01}}
	
\end{enumerate}

This construction implies the `reproducing property' of the positive-definitive kernel that gives rise to $ \mathcal{H} $: 
\begin{equation*}
\langle k(\cdot,x), f \rangle = f(x),
\end{equation*}
see also \textcite[][Section 2.2., p.33]{kernels01} for more details.
%\ccomment{Wir koennten einfach damit anfangen und sagen, dass sowas exisitiert und auf Literatur oder Appendix verweisen? Sehe gerade auch nicht, wie das aus 1.-4. folgt}\pcomment{die referenz für diese herleitung ist \textcite[][Section 2.2., p.33]{kernels01}}
In particular, we obtain
\begin{equation*}
\langle k(\cdot,x), k(\cdot,x') \rangle = \langle \Phi(x),\Phi(x')\rangle = k(x,x').
\end{equation*}

%\newpage

\section{Further Simulation Results}\label{app_more_results}

In this section, we present further simulations results analogous to Figures \ref{fig_mc_var1_causal}, \ref{fig_mc_var1_anticausal}, \ref{fig_mc_var1_causal_discovery} but with  $ \mathrm{Var}(U) = 0.8 $ and $ \mathrm{Var}(U) = 1.2 $. 
%The results are not sensitive to changes in the error variance $ \mathrm{Var}(U)$.

\begin{figure}[t]
	\centering
	\includegraphics[width=1\textwidth]{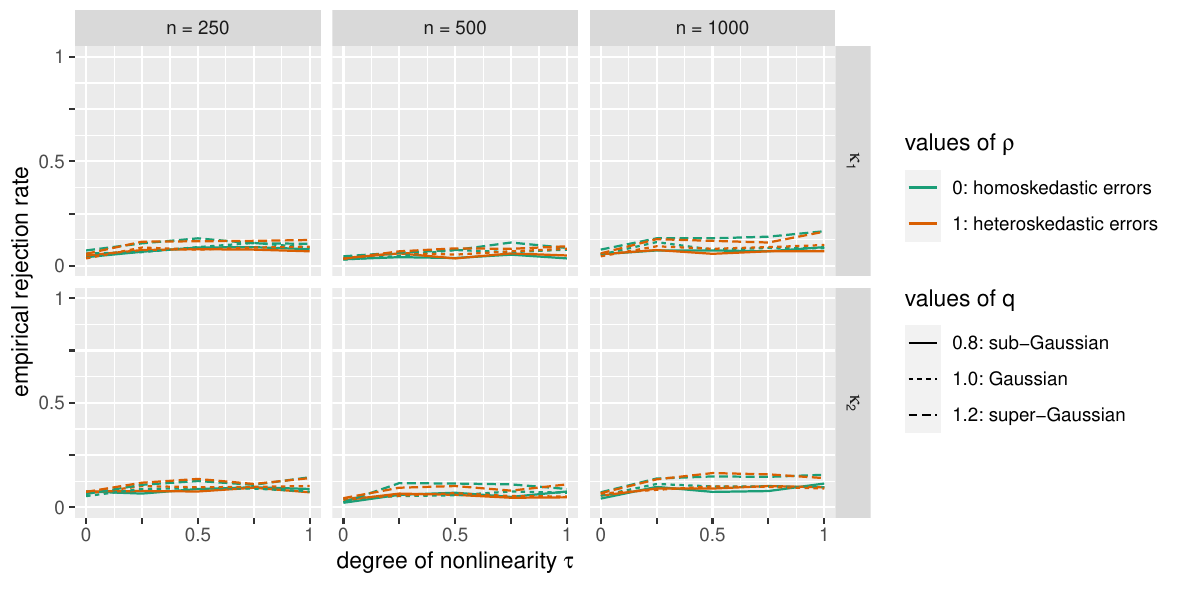}
	\caption{
		Empirical rejection rates of independence of estimated residual and $X$ of model \eqref{eq_w_causal} when $c_{\rho,q}$ is chosen such that $\mathrm{Var}(U) = .8$. 
	}
	\label{fig_mc_var89_causal}
\end{figure}

\begin{figure}[t]
	\centering
	\includegraphics[width=1\textwidth]{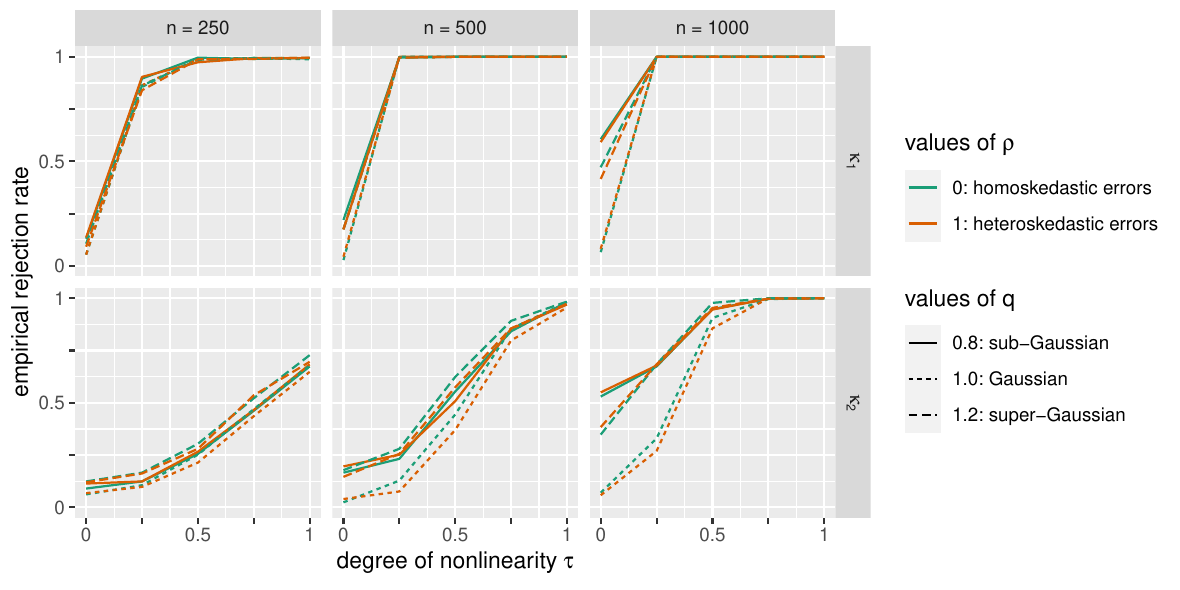}
	\caption{
		Empirical rejection rates of independence of estimated residual and $Y$ of model \eqref{eq_w_backw1} when $c_{\rho,q}$ is chosen such that $\mathrm{Var}(U) = .8$. 
	}
	\label{fig_mc_var89_anticausal}
\end{figure}

\begin{figure}[t]
	\centering
	\includegraphics[width=1\textwidth]{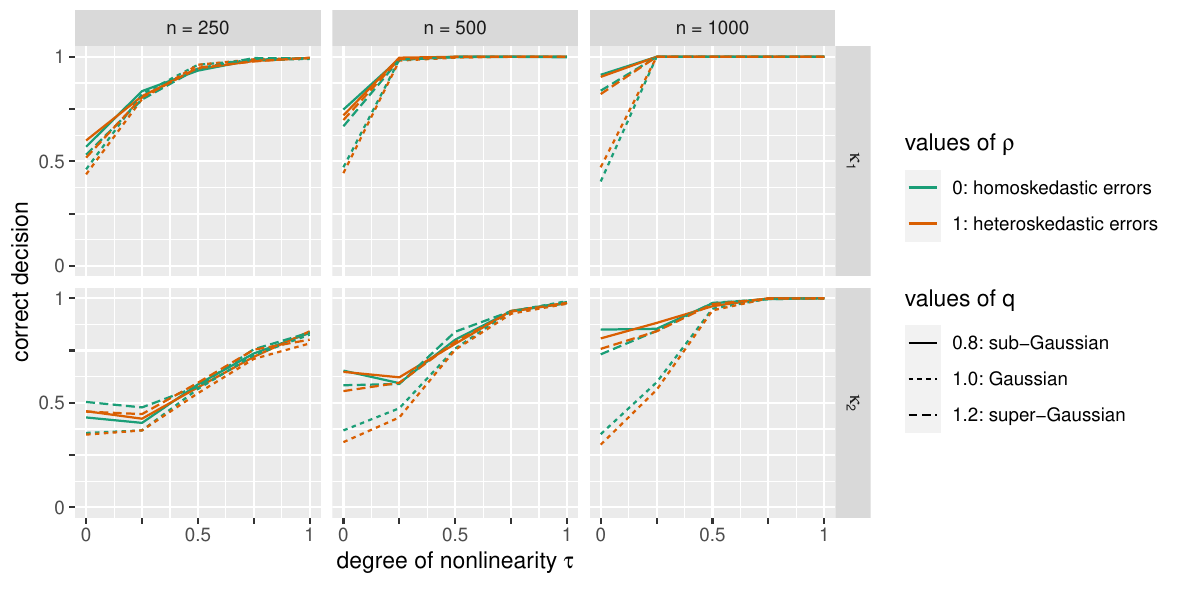}
	\caption{
		Empirical probabilities of correct recovery of the causal direction when $c_{\rho,q}$ is chosen such that $\mathrm{Var}(U) = .8$. 
	}
	\label{fig_mc_var89_causal_discovery}
\end{figure}

\begin{figure}[t]
	\centering
	\includegraphics[width=1\textwidth]{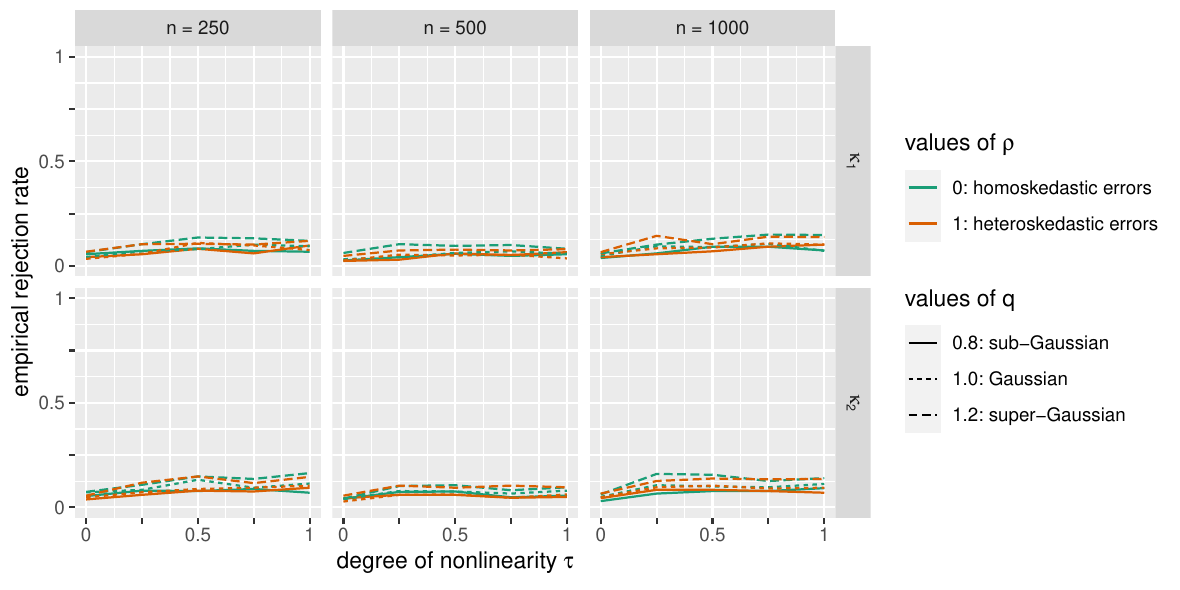}
	\caption{
		Empirical rejection rates of independence of estimated residual and $X$ of model \eqref{eq_w_causal} when $c_{\rho,q}$ is chosen such that $\mathrm{Var}(U) = 1.2$. 
	}
	\label{fig_mc_var11_causal}
\end{figure}

\begin{figure}[t]
	\centering
	\includegraphics[width=1\textwidth]{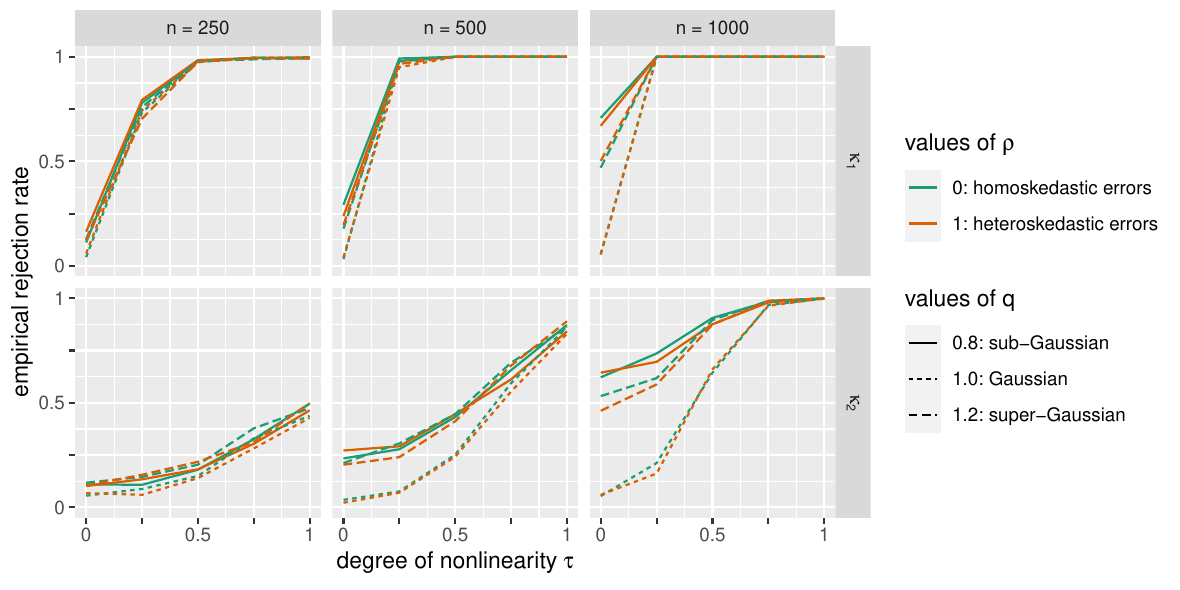}
	\caption{
		Empirical rejection rates of independence of estimated residual and $Y$ of model \eqref{eq_w_backw1} when $c_{\rho,q}$ is chosen such that $\mathrm{Var}(U) = 1.2$. 
	}
	\label{fig_mc_var11_anticausal}
\end{figure}

\begin{figure}[t]
	\centering
	\includegraphics[width=1\textwidth]{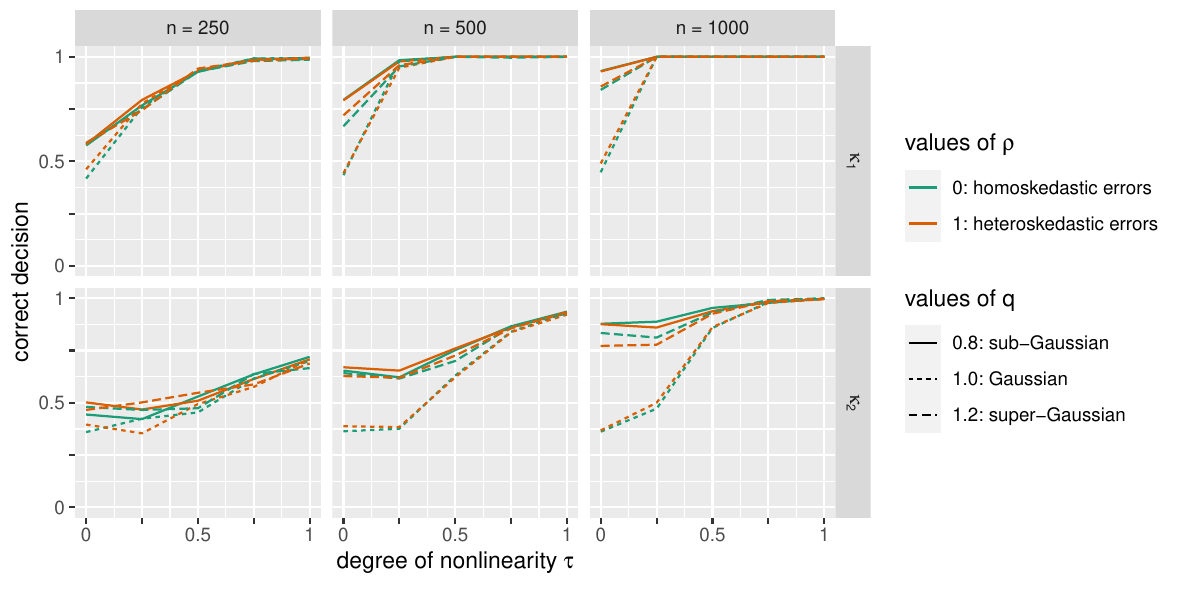}
	\caption{
		Empirical probabilities of correct recovery of the causal direction when $c_{\rho,q}$ is chosen such that $\mathrm{Var}(U) = 1.2$. 
	}
	\label{fig_mc_var11_causal_discovery}
\end{figure}
\end{document}